\documentclass{article}
\usepackage{eurosym}
\usepackage{amssymb}
\usepackage{amsmath}
\usepackage{amsfonts}
\usepackage{version}
\usepackage{appendix}
\usepackage{portland}
\usepackage{lscape}
\usepackage{setspace}

\input{tcilatex}
\begin{document}

\title{Modelling Returns and Volatilities During Financial Crises: a Time
Varying Coefficient Approach }
\author{M. Karanasos$^{\dagger ^{\ast }}$, A. G. Paraskevopoulos$^{\ddagger
} $, F.Menla Ali$^{\dagger }$, M. Karoglou$^{\star }$, S. Yfanti$^{\dagger }$
\\
$^{\dagger }$\textit{Economics and Finance}, \textit{Brunel University,
London, UK}\\
$^{\ddagger }$\textit{CRANS}, \textit{Mathematics}, \textit{University of
Patras, Patra, Greece}\\
$^{\star }$\textit{Aston Business School, Birmingham UK}}
\date{This draft: March 27th 2014\\
}
\maketitle

\begin{abstract}
We examine how the most prevalent stochastic properties of key financial
time series have been affected during the recent financial crises. In
particular we focus on changes associated with the remarkable economic
events of the last two decades in the mean and volatility dynamics,
including the underlying volatility persistence and volatility spillovers
structure. Using daily data from several key stock market indices we find
that stock market returns exhibit time varying persistence in their
corresponding conditional variances. Furthermore, the results of our
bivariate GARCH models show the existence of time varying correlations as
well as time varying shock and volatility spillovers between the returns of
FTSE and DAX, and those of NIKKEI and Hang Seng, which became more prominent
during the recent financial crisis. Our theoretical considerations on the
time varying model which provides the platform upon which we integrate our
multifaceted empirical approaches are also of independent interest. In
particular, we provide the general solution for low order time varying
specifications, which is a long standing research topic. This enables us to
characterize these models by deriving, first, their multistep ahead
predictors, second, the first two time varying unconditional moments, and
third, their covariance structure.



{\footnotesize We gratefully} {\footnotesize acknowledge the helpful
comments of a referee and the helpful correspondence we had with L.
Giraitis, G. Kapetanios and A. Magdalinos in the preparation of this paper.
We would also like to thank R. Baillie, L. Bauwens, M. Brennan, D. van Dijk,
W. Distaso, C. Francq, C. Gourieroux, E. Guerre, M. Guidolin, A. Harvey, C.
Hommes, S. Leybourne, P. Minford, A. Monfort, C. Robotti, W. Semmler, R.
Smith, T. Ter\"{a}svirta, P. Zaffaroni, and J-M Zakoian for suggestions and
comments on a closely related work (see Paraskevopoulos et al., 2013), which
greatly improved many aspects of the current paper as well. We are grateful
to seminar participants at CREST, Erasmus University, London School of
Economics, Queen Mary University of London, Imperial College, University of
Essex, Birkbeck College University of London, University of Nottingham,
Cardiff University, University of Manchester, Athens University of Economics
and Business, and University of Piraeus. We have also benefited from the
comments given by participants (on the closely related work) at the 3rd
Humboldt-Copenhagen Conference on Financial Econometrics (Humboldt
University, Berlin, March 2013), the} {\footnotesize SNDE 21st Annual
Symposium (University of Milan-Bicocca, March 2013), the 8th and 9th
BMRC-QASS conferences on Macro and Financial Economics (Brunel University,
London, May 2013), the 7th CFE Conference (Senate House, University of
London, December 2013), and the 1st RASTANEWS Conference (University of
Milan-Bicocca, January 2014).}

$^{\ast }${\footnotesize Address for correspondence: Menelaos Karanasos,
Economics and Finance, Brunel University, West London, UB3 3PH, UK; email:
menelaos.karanasos@brunel.ac.uk, tel: +44(0)1895265284, fax: +44
(0)1895269770.}

$^{^{\ast }}${\footnotesize The order of the authors' names reflect their
contribution. M. Karanasos and A. Paraskevopoulos are joint first authors,
having defined the theoretical and empirical models, and having derived the
theoretical results (Sections 2 and 3). F.\ Menla Ali is second author,
having estimated the various univariate and bivariate models in Sections 5
and 6. M. Karoglou, by applying Karoglou (2010) tests, incorporated the
break detection procedure into the empirical analysis (Section 5.1). S.
Yfanti derived the autocorrelations in Section 3.1.2 and the unconditional
variances in Section 3.2.1.}
\end{abstract}

\ \newpage

\section{Introduction}

\doublespacing

The global financial crisis of 2007-08 and the European sovereign-debt
crisis that took place immediately afterwards are at the heart of the
research interests of practitioners, academics, and policy makers alike.
Given the widespread fear of an international systemic financial collapse at
the time it is no wonder that the currently on-going heated discussion on
the actual causes and effects of these crises is the precursor to the
development of the necessary tools and policies for dealing with similar
phenomena in the future.

The inevitable step in undertaking such an enormous task is to map, as
accurately as possible, the `impact' of these crises onto what are currently
considered the main stochastic properties of the underlying financial time
series. In this way, informed discussions on the causes and effects of these
crises can take place and thus more accurately specify the set of features
that have to characterize the necessary tools and policies to address them.
This paper aspires to provide a platform upon which changes in the main
statistical properties of financial time series due to economic crises can
be measured.

In particular, we focus on the recent financial crises and examine how the
mean and volatility dynamics, including the underlying volatility
persistence and volatility spillovers structure, have been affected by these
crises. With this aim we make use of several modern econometric approaches
for univariate and multivariate time series modelling, which we also
condition on the possibility of breaks in the mean and/or volatility
dynamics taking place. Moreover, we unify these approaches by introducing a
set of theoretical considerations for time varying (TV) AR-GARCH models,
which are also of independent interest. In particular, we make three broad
contributions to the existing literature.

First, we present and utilize some new theoretical results on time varying
AR and/or asymmetric GARCH (AGARCH) models. 
We limit our analysis to low order specifications to save space
and also since it is well \ documented that low order AR models for stock
returns often emerge in practice. We show the applicability of these general
results to one important case: that of abrupt breaks, which we make
particular use of in our empirical investigation. Our models produce time
varying unconditional variances in the spirit of Engle and Rangel\ (2008)
and Baillie and Morana (2009). TV-GARCH specifications have recently gained
popularity for modelling structural breaks in the volatility process (see,
for example, Frijns et al., 2011, and Bauwens, et al., 2014). Despite nearly
half a century of research work and the widely recognized importance of time
varying models, until recently there was a lack of a general theory that can
be employed to explore their time series properties systematically. Granger
in some of his last contributions highlighted the importance of the topic
(see, Granger 2007, and 2008). The stumbling block to the development of
such a theory was the lack of a method that can be used to solve time
varying difference equations of order two or higher. Paraskevopoulos et al.
(2013) have developed such a general theory (see also Paraskevopoulos and
Karanasos, 2013). The starting point of the solution method that we present
below is to represent the linear time varying difference equation of order
two as an infinite system of linear equations. The coefficient matrix of
such an infinite system is row finite. The solution to such infinite systems
is based on an extension of the classic Gauss elimination, called Infinite
Gaussian elimination (see Paraskevopoulos, 2012, 2014). Our method is a
natural extension of the first order solution formula. It also includes the
linear difference equation with constant coefficients (see, for example,
Karanasos, 1998, 2001) as a special case. We simultaneously compute not only
the general solution but its homogeneous and particular parts as well. The
coefficients in these solutions are expressed as determinants of tridiagonal
matrices. This allows us to provide a thorough description of time varying
models by deriving, first, multistep ahead forecasts, the associated
forecast error and the mean square error, and second, the first two time
varying unconditional moments of the process and its covariance structure.

Second, we use a battery of tests to identify the number and estimate the
timing of breaks both in the mean and volatility dynamics. Following our
theoretical results and prompted by Morana and Beltratti (2004) amongst
others who acknowledge that misleading inference on the persistence of the
volatility process may be caused by unaccounted structural breaks\footnote{%
A detailed literature review on this issue is available upon request.}, we
implement these break tests in the univariate context also to determine
changes in the persistence of volatility. The special attention we pay to
this issue is well justified, especially within the finance literature given
that it is well-established that the proper detection of breaks is pivotal
for a variety of financial applications, particularly in risk measurement,
asset allocation and option pricing. Kim and Kon (1999) emphasize the
importance of incorporating some break detection procedure into the existing
financial modelling paradigms when they call attention to the fact that
"\ldots \textit{Public announcements of corporate investment and financial
decisions that imply a change in the firm's expected return and risk will be
impounded in stock prices immediately in an efficient market. The
announcements of relevant macroeconomic information will affect the return
and risk of all securities, and hence, portfolios (indexes). Since relevant
information that changes the risk structure is randomly released with some
time interval (not at every moment) in sequence, these information events
translate into sequential discrete structural shifts (or change-points) for
the mean and/or variance parameter(s) in the time series of security returns.%
}"

Third, we employ the bivariate unrestricted extended dynamic conditional
correlation (UEDCC) AGARCH process to analyze the volatility transmission
structure, applied to stock market returns. The model is based on the
dynamic conditional correlation of Engle (2002) allowing for volatility
spillovers effects by imposing the unrestricted extended conditional
correlation (dynamic or constant) GARCH specification of Conrad and
Karanasos (2010). The most recent applications of the model can be found in
Conrad et al. (2010), Rittler (2012), Karanasos and Zeng (2013) and Conrad
and Karanasos (2013). However, we extend it by allowing shock and volatility
spillovers parameters to shift across abrupt breaks as well as across two
regimes of stock returns, positive (increases in the stock market) and
negative (declines in the stock market) (see also Karanasos et al., 2013).
Recently, following our work, Caporale et al. (2014) adopted our UEDCC
framework but they do not allow for breaks in the shock and volatility
spillovers. \ The extant literature on modelling returns and volatilities is
extensive, and it has evolved in several directions. One line of literature
has focused on return correlations 
and comovements or what is known as contagion among different
markets or assets (e.g., Caporale et. al., 2005; Rodriguez, 2007, among
others), while another line of the literature has focused on volatility
spillovers among the markets (e.g., Baele, 2005; Asgharian and Nossman,
2011, among others). The model adopted in this paper is flexible enough to
capture contagion effects as well as to identify the volatility spillovers
associated with the structural changes and exact movements of each market
(e.g., upward or downward) to the other, and vice versa. Knowledge of this
mechanism can provide important insights to investors by focusing their
attention on structural changes in the markets as well as their trends and
movements (e.g., upward or downward) in order to set appropriate portfolio
management strategies.

Overall, our results suggest that stock market returns exhibit time varying
persistence in their corresponding conditional variances. The results of the
bivariate UEDCC-AGARCH($1,1$) model applied to FTSE and DAX returns, and to
NIKKEI and Hang Seng returns, show the existence of dynamic correlations as
well as time varying shock and volatility spillovers between the two
variables in each pair. For example, the results of the bivariate FTSE and
DAX returns show that the transmission of volatility from DAX to FTSE
exhibited a time varying pattern across the Asian financial crisis and the
announcement of the \euro 18bn German tax cuts plan as well as the global
financial crisis. As far as the NIKKEI and Hang Seng pair is concerned, the
results provide evidence that these two financial markets have only been
integrated during the different phases of the recent financial crisis. With
regard to the regime-dependent volatility spillovers, the results suggest
that declines in FTSE and DAX generate shock spillovers to each other,
whereas increases in each of these market generate negative volatility
spillovers to the other. Furthermore, the results show that declines in
NIKKEI generate shock spillovers to Hang Seng, whilst increases in NIKKEI
generate negative volatility spillovers to Hang Seng.

The remainder of this paper is as follows. Section 2 considers the AR-GARCH
model with abrupt breaks in the first two conditional moments, and the time
varying process, which are our two main objects of inquiry. Section 3
introduces the theoretical considerations on the time varying AR and AGARCH
models. In Section 3.1 we represent the former as an infinite linear system
and concentrate on the associated coefficient matrix. This representation
enables us to establish an explicit formula for the general solution in
terms of the determinants of tridiagonal matrices. We also obtain the
statistical properties of the aforementioned models, e.g., multi-step-ahead
predictors and their forecast error variances. Section 4 describes our
methodology and data. Section 5 presents our empirical univariate results,
and the next Section discusses the results from various bivariate models.
The final Section contains the summary and our concluding remarks.

\section{Abrupt Breaks}

First, we introduce the notation and the AR-AGARCH model with abrupt breaks
both in the conditional mean and variance. Throughout the paper we will
adhere to the conventions: ($\mathbb{Z}^{+}$) $\mathbb{Z},$ and ($\mathbb{R}%
^{+}$) $\mathbb{R}$ stand for the sets of (positive) integers, and
(positive) real numbers, respectively. To simplify our exposition we also
introduce the following notation. Let $t\in $ $\mathbb{Z}$ represents
present time, and $k\in \mathbb{Z}^{+}$ the prediction horizon.

\subsection{The Conditional Mean}

In this paper we will examine an AR($2$) model\footnote{%
To keep the exposition tractable and reveal its practical significance we
work with low order specifications.} with $n$ abrupt breaks, $0\leq n\leq
k-1 $, at times $t-k_{1}$, $t-k_{2}$, $\ldots $, $t-k_{n}$, where $%
0=k_{0}<k_{1}<k_{2}<\cdots <k_{n}<k_{n+1}=k$, $k_{l}\in \mathbb{Z}^{+}$, and 
$k_{n}$ is finite. That is, between $t-k=t-k_{n+1}$\ and the present time $%
t=t-k_{0}$ the AR process contains $n$ structural breaks and the switch from
one set of parameters to another is abrupt. In particular,%
\begin{equation}
y_{\tau }=\phi _{0,l}+\phi _{1,l}y_{\tau -1}+\phi _{2,l}y_{\tau
-2}+\varepsilon _{\tau },  \label{ASBAR(2)}
\end{equation}%
for $l=1,\ldots ,n+1,$\ and $\tau =t-k_{l-1},\ldots ,t-k_{l}+1$, where%
\footnote{$\{ \tciFourier _{t}\}$\ is a non-decreasing sequence of $\sigma $%
-fields $\tciFourier _{t-1}\subseteq \tciFourier _{t}\subseteq \tciFourier $.%
} $\mathbb{E}[\varepsilon _{\tau }\left \vert \tciFourier _{\tau -1}\right.
]=0$ and $\varepsilon _{\tau }$ follows a time varying AGARCH type of
process with finite variance $\sigma _{\tau }^{2}$ (see the next Section).%
\footnote{%
Without loss of generality we will assume that outside the prediction
horizon there are no breaks. That is: regime one ($l=1$) extends to time $%
\tau =\ldots ,t+2,t+1$ and the ($n+1$)th regime extends to time $\tau
=t-k,t-k-1,\ldots $.} Within the class of AR($2$) processes, this
specification is quite general and allows for intercept and slope shifts as
well as errors with time varying variances (see also Pesaran and Timmermann,
2005, and Pesaran et al. 2006). Each regime $l$\ is characterized by a
vector of regression coefficients, $\mathbf{\phi }_{l}=(\phi _{0,l},\phi
_{1,l},\phi _{2,l})^{\prime }$, and positive and finite time varying
variances, $\sigma _{\tau }^{2}$, $\tau =t-k_{l-1},\ldots ,t-k_{l}+1$. We
will term the AR($2$) model with $n$ abrupt breaks: abrupt breaks
autoregressive process of order ($2;n$), AB-AR($2;n$).

\subsection{The Conditional Variance\label{SubsecVariance1}}

We assume that the noise term is characterized by the relation $\varepsilon
_{\tau }=e_{\tau }\sqrt{h_{\tau }}$, where $h_{\tau }$ is positive with
probability one and it is a measurable function of $\mathcal{F}_{t-1}$; $%
e_{\tau }$ is an i.i.d sequence with zero mean and finite second and fourth
moments: $\varkappa ^{(i)}=\mathbb{E}(e_{\tau }^{2i})$, $i=1,2$. In other
words the conditional (on time $\tau -1$) variance of $y_{\tau }$ is $%
\mathbb{V}ar(y_{\tau }\left \vert \mathcal{F}_{\tau -1}\right. )=\varkappa
^{(1)}h_{\tau }$. In what follows, without loss of generality, we will
assume that $\varkappa ^{(1)}=1$.

Moreover, we specify the parametric structure of $h_{\tau }$ as an AGARCH($%
1,1$) model with $m$ abrupt breaks, $0\leq m\leq k-1$, at times $t-\kappa
_{1}$, $t-\kappa _{2}$, $\ldots $, $t-\kappa _{m}$, where $0=\kappa
_{0}<\kappa _{1}<\kappa _{2}<\cdots <\kappa _{m}<\kappa _{m+1}=k$, $\kappa
_{m}\in \mathbb{Z}^{+}$, and $\kappa _{m}$ is finite. That is, between $%
t-k=t-\kappa _{m+1}$\ and the present time $t=t-\kappa _{0}$ the AGARCH
process contains $m$ structural breaks and the switch from one set of
parameters to another is abrupt: 
\begin{equation}
h_{\tau }=\omega _{\ell }+\alpha _{\ell }^{\ast }\varepsilon _{\tau
-1}^{2}+\beta _{\ell }h_{\tau -1},  \label{ABR-AGARCH(1,1)}
\end{equation}%
for $\ell =1,\ldots ,m+1,$\ and $\tau =t-\kappa _{\ell -1},\ldots ,t-\kappa
_{\ell }+1$; where $\alpha _{\ell }^{\ast }\triangleq \alpha _{\ell }+\gamma
_{\ell }S_{\tau -1}^{-}$, with $S_{\tau -1}^{-}=1$ if $e_{\tau -1}<0$, $0$
otherwise.\footnote{%
This type of asymmetry is the so called GJR-GARCH model (named for Glosten
et al., 1993). The asymmetric power ARCH process (see, among others,
Karanasos and Kim, 2006; Margaronis et al., 2013) is yet another asymmetric
variant. For other asymmetric GARCH models see Francq and Zako\"{\i}an
(2010, chapter 10) and the references therein.} \ As with the AR process we
will assume that outside the prediction horizon there are no breaks.
Obviously, the above process nests the simple AGARCH($1,1$) specification if
we assume that the four coefficients are constant.

In what follows we provide a complete characterization of the main
time-series properties of this model. Although in this work we will focus
our attention on the AB-AR($2;n$)-AGARCH($1,1;m$) process\footnote{%
That is an AR($2$)-AGARCH($1,1$) model with $n$ and $m$ abrupt breaks in the
conditional mean and variance respectively.} our results can easily be
extended to models of higher orders (see Paraskevopoulos et al., 2013).

\subsection{Time Varying Model}

In the current Section we face the non-stationarity of processes with abrupt
breaks head on by employing a time varying treatment. In particular, we put
forward a framework for examining the AR-AGARCH specification with $n$ and $%
m $ abrupt breaks in the conditional mean and variance respectively. We
begin by expressing the model as a TV-AR($2$)-AGARCH($1,1$) process:%
\begin{equation}
y_{t}=\phi _{0}(t)+\phi _{1}(t)y_{t-1}+\phi _{2}(t)y_{t-2}+\varepsilon _{t},
\label{TVAR(2)}
\end{equation}%
where for $l=1,\ldots ,n+1$ and $\tau =t-k_{l-1},\ldots ,t-k_{l}+1$, $\phi
_{i}(\tau )\triangleq \phi _{i,l}$, $i=0,1,2$, are the time varying drift
and AR parameters; as before $\{ \varepsilon _{t},t\in 
\mathbb{Z}
\}$ is a sequence of zero mean serially uncorrelated random variables with
positive and finite time varying variances $\sigma _{t}^{2}$ $\forall $ $t$. 
Recall that we have relaxed the assumption of homoscedasticity
that is likely to be violated in practice and allow $\varepsilon _{t}$ to
follow a TV-AGARCH($1,1$) type of process:%
\begin{equation}
h_{t}=\omega (t)+\alpha ^{\ast }(t)\varepsilon _{t-1}^{2}+\beta (t)h_{t-1},
\label{TVGARCH}
\end{equation}%
where for $\ell =1,\ldots ,m+1$ and $\tau =t-\kappa _{\ell -1},\ldots
,t-\kappa _{\ell }+1$, $\omega (\tau )\triangleq \omega _{\ell }$, $\alpha
^{\ast }(\tau )\triangleq \alpha (\tau )+\gamma (\tau )S_{t-1}^{-}\triangleq
\alpha _{\ell }^{\ast }$, and $\beta (\tau )\triangleq \beta _{\ell }$ are
the time varying parameters of the conditional variance equation.

The TV-AGARCH($1,1$) formulation in eq. (\ref{TVGARCH}) can readily be seen
to have the following representation%
\begin{equation}
h_{t}=\omega (t)+c(t)h_{t-1}+\alpha ^{\ast }(t)v_{t-1},  \label{TV-ARMA(2,2)}
\end{equation}%
with $c(t)\triangleq \alpha ^{\ast }(t)+\beta (t)=\alpha (t)+\gamma
(t)S_{t-1}^{-}+\beta (t)$, and for $\ell =1,\ldots ,m+1$ and $\tau =t-\kappa
_{\ell -1},\ldots ,t-\kappa _{\ell }+1$, $c(\tau )\triangleq c_{\ell }$; the
`innovation' of the conditional variance $v_{t}=\varepsilon _{t}^{2}-h_{t}$
is, by construction, an uncorrelated term with expected value $0$ and $%
\mathbb{E}(v_{t}^{2})=\sigma _{vt}^{2}=\widetilde{\varkappa }\mathbb{E}%
(h_{t}^{2})$ (the conditions for the second unconditional moments, $\mathbb{E%
}(h_{t}^{2})$, to exist for all $t$\ are available upon request), with $%
\widetilde{\varkappa }=\mathbb{V}ar(e_{t}^{2})=\varkappa ^{(2)}-1$. The
above equation has the linear structure of a TV-ARMA model allowing for
simple computations of the linear predictions (see Section \ref%
{SubsubsecErrorVariance} below).\footnote{%
As pointed out, among others, by Francq and Zako\"{\i}an (2010, p. 20) under
additional assumptions (implying the second-order of $h_{t}$ or $\varepsilon
_{t}^{2}$), which in our case are available upon request, we can state that
if $\varepsilon _{t}$ follows a TV-AGARCH model then $h_{t}$ or $\varepsilon
_{t}^{2}$ are TV-ARMA processes as well.}

Although in the next Section we will focus our attention on the TV-AR($2$%
)-AGARCH($1,1$) model our results can easily be extended to time varying
models of higher orders (see Paraskevopoulos et al., 2013).

\section{Theoretical Considerations}

This current Section presents some new theoretical findings for time varying
models which also provide the platform upon which we unify the results we
obtain from the different econometric tools. \ That is, we put forward a
framework for examining AR models with abrupt breaks, like eq.(\ref{ASBAR(2)}%
), based on a workable closed form solution of stochastic time varying
difference equations. In other words, we exemplify how our theoretical
methodology can be used to incorporate structural changes, which in this
paper we view as abrupt breaks. \ We also explain how we can extend our
approach to the AGARCH specification with abrupt breaks in the conditional
variance.

\subsection{The mean\label{SubsecMeanTheorConsid}}

In the context of eq. (\ref{TVAR(2)}), the second-order homogeneous
difference equation with time varying coefficients is written as%
\begin{equation}
\phi _{2}(t)y_{t-2}+\phi _{1}(t)y_{t-1}-y_{t}=0,\text{ }t\geq \tau +1=t-k+1.
\label{Difference2}
\end{equation}%
The infinite set of equations in the above equation is equivalent to the
infinite linear system whose coefficient matrix is row-finite (row-finite
matrices are infinite $\mathbb{N}\times \mathbb{N}$ matrices whose rows have
a finite number of nonzero elements)

\begin{equation}
\left( 
\begin{array}{ccccccc}
\phi _{2}(\tau +1) & \phi _{1}(\tau +1) & -1 &  &  &  & \cdots \\ 
& \phi _{2}(\tau +2) & \phi _{1}(\tau +2) & -1 &  &  & \cdots \\ 
&  & \phi _{2}(\tau +3) & \phi _{1}(\tau +3) & -1 &  & \cdots \\ 
\vdots & \vdots & \vdots & \vdots & \vdots & \vdots & \vdots \vdots \vdots%
\end{array}%
\right) \left( 
\begin{array}{l}
y_{\tau -1} \\ 
y_{\tau } \\ 
y_{\tau +1} \\ 
y_{\tau +2} \\ 
y_{\tau +3} \\ 
y_{\tau +4} \\ 
\vdots%
\end{array}%
\right) =\left( 
\begin{array}{l}
0 \\ 
0 \\ 
0 \\ 
\vdots%
\end{array}%
\right) ,  \label{PHIMatrix}
\end{equation}
(here and in what follows empty spaces in a matrix\footnote{%
Matrices and vectors are denoted by upper and lower case boldface symbols,
respectively. For square matrices $\mathbf{X}=[x_{ij}]_{i,j=1,\ldots ,k}\in 
\mathbb{R}^{k\times k}$ using standard notation, det$(\mathbf{X})$ or $%
\left
\vert \mathbf{X}\right \vert $ denotes the determinant of matrix $%
\mathbf{X}$.} have to be replaced by zeros) or in a compact form: $\mathbf{%
\Phi \cdot y}=\mathbf{0}$\textbf{.} The equivalence of eqs. (\ref%
{Difference2}) and (\ref{PHIMatrix}) follows from the fact that for an
arbitrary $i$ in $\{1,2,3,\ldots \}$ the $i$th equation of (\ref{PHIMatrix}%
), as a result of the multiplication of the $i$th row of $\mathbf{\Phi }$ by
the column of $y${\footnotesize s} equated to zero, is equivalent to eq. (%
\ref{Difference2}), as of time $t=\tau +i$. By deleting the first column of
the $\mathbf{\Phi }$ matrix and then keeping only the first $k$ rows and
columns we obtain the following square matrix:

\begin{eqnarray}
\mathbf{\Phi }_{t,k} &=&\left( 
\begin{array}{cccccc}
\phi _{1}(\tau +1) & -1 &  &  &  &  \\ 
\phi _{2}(\tau +2) & \phi _{1}(\tau +2) & -1 &  &  &  \\ 
& \phi _{2}(\tau +3) & \phi _{1}(\tau +3) & -1 &  &  \\ 
&  & \ddots & \ddots & \ddots &  \\ 
&  &  & \phi _{2}(t-1) & \phi _{1}(t-1) & -1 \\ 
&  &  &  & \phi _{2}(t) & \phi _{1}(t)%
\end{array}%
\right)  \notag \\
&&  \label{FAIAR(2)}
\end{eqnarray}%
(where $\tau =t-k$). Formally $\mathbf{\Phi }_{t,k}$ is a square $k\times k$
matrix whose ($i,j$) entry $1\leq i,j\leq k$ is given by

\begin{equation*}
\left \{ 
\begin{array}{cccccc}
-1 & \text{if} &  & i=j-1, & \text{and} & 2\leq j\leq k, \\ 
\phi _{1+d}(t-k+i) & \text{if} & d=0,1, & i=j+d, & \text{and} & 1\leq j\leq
k-d, \\ 
0 &  & \text{otherwise.} &  &  & 
\end{array}%
\right.
\end{equation*}

This is a tridiagonal or continuant matrix, that is a matrix that is both
upper and lower Hessenberg matrix. We next define the bivariate function $%
\xi :\mathbb{Z}\times \mathbb{Z}^{+}\longmapsto \mathbb{R}$ by 
\begin{equation}
\xi _{t,k}=\text{det}(\mathbf{\Phi }_{t,k})  \label{KSIAR(2)}
\end{equation}%
coupled with the initial values $\xi _{t,0}=1$, and $\xi _{t,-1}=0$. $\xi
_{t,k}$ for $k\geq 2$, is a determinant of a $k\times k$ matrix; each two
nonzero diagonals (below the superdiagonal) of this matrix consists of the
time varying coefficients $\phi _{i}(\mathfrak{\cdot })$, $i=1,2$, from $%
t-k+i$ to $t$. That is, the number of elements of $\phi _{i}(\mathfrak{\cdot
)}$ in the diagonals below the superdiagonal is $k-i+1$. In other words, $%
\xi _{t,k}$ is a\ $k$-order tridiagonal determinant. For the AB-AR($2;n$)
process, $\xi _{t,k}$ is given by 
\begin{equation}
\xi _{t,k}=\det (\mathbf{\Phi }_{t,k})=\left \vert 
\begin{array}{cccccccc}
\phi _{1,n+1} & -1 &  &  &  &  &  &  \\ 
\phi _{2,n+1} & \phi _{1,n+1} & -1 &  &  &  &  &  \\ 
& \ddots & \ddots & \ddots &  &  &  &  \\ 
&  & \phi _{2,l} & \phi _{1,l} & -1 &  &  &  \\ 
&  &  & \phi _{2,l} & \phi _{1,l} & -1 &  &  \\ 
&  &  &  & \ddots & \ddots & \ddots &  \\ 
&  &  &  &  & \phi _{2,1} & \phi _{1,1} & -1 \\ 
&  &  &  &  &  & \phi _{2,1} & \phi _{1,1}%
\end{array}%
\right \vert ,  \label{KSIABAR(2)}
\end{equation}%
that is, the ($i,i-1$), and ($i,i$) elements in rows $i=k-k_{l-1},\ldots
,k-(k_{l}-1)$, $l=1,\ldots ,n+1$, of the matrix $\mathbf{\Phi }_{t,k}$ are
given by $\phi _{2,l}$ and $\phi _{1,l}$, respectively. 

The general term of the general homogeneous solution of eq. (\ref%
{Difference2}) with two free constants (initial condition values), $y_{t-k}$
and $y_{t-k-1}$, is given by%
\begin{equation}
y_{t,k}^{hom}=\xi _{t,k}y_{t-k}+\phi _{2}(t-k+1)\xi _{t,k-1}y_{t-k-1}.
\label{TVAR2HomSol}
\end{equation}

Similarly, the general particular solution, $y_{t,k}^{par}$, can be
expressed as%
\begin{equation}
y_{t,k}^{par}=\sum_{r=0}^{k-1}\xi _{t,r}\phi _{0}(t-r)+\sum_{r=0}^{k-1}\xi
_{t,r}\varepsilon _{t-r}.  \label{ParticularSolution}
\end{equation}

The general solution of eq. (\ref{TVAR(2)}) with free parameters $y_{t-k}$, $%
y_{t-k-1}$ is given by the sum of the homogeneous solution plus the
particular solution: 
\begin{equation}
y_{t,k}^{gen}=y_{t,k}^{hom}+y_{t,k}^{par}=\xi _{t,k}y_{t-k}+\phi
_{2}(t-k+1)\xi _{t,k-1}y_{t-k-1}+\sum_{r=0}^{k-1}\xi _{t,r}\phi
_{0}(t-r)+\sum_{r=0}^{k-1}\xi _{t,r}\varepsilon _{t-r}.  \label{TVAR(2)SOL}
\end{equation}%
(see the Appendix and also Paraskevopoulos et al., 2013 and Karanasos et
al., 2014a). \ In the above expression $y_{t,k}^{gen}$\ is decomposed into
two parts: the $y_{t,k}^{hom}$\ part, which is written in terms of the two\
free constants ($y_{t-k-i}$, $i=0,1$); and, the $y_{t,k}^{par}$\ part, which
contains the\ time varying drift terms ($\phi _{0}(\mathfrak{\cdot })$) and
the error terms ($\varepsilon ${\footnotesize s}) from time $t-k+1$\ to time 
$t$. When $k=1$, since $\xi _{t,0}=1$ and $\xi _{t,1}=\phi _{1}(t)$, the
above expression reduces to eq. (\ref{TVAR(2)}). Notice also that for the
model with $n$ abrupt breaks, we have 
\begin{equation*}
\sum_{r=0}^{k-1}\xi _{t,r}\phi _{0}(t-r)=\sum_{l=1}^{n+1}\phi
_{0,l}\sum_{r=k_{l-1}}^{k_{l}-1}\xi _{t,r},\text{ and }\phi _{2}(t-k+1)=\phi
_{2,n+1},
\end{equation*}%
where $\xi _{t,r}$ is given in eq. (\ref{KSIABAR(2)}). The main advantage of
our TV model/methodology is that we suppose that the law of evolution of the
parameters is unknown, in particular they may be stochastic (i.e., we can
either have a stationary or non-stationary process) or non stochastic (e.g.,
periodic models serve as an example, see Karanasos et al., 2014a,b).
Therefore, no restrictions are imposed on the functional forms of the time
varying AR parameters. In the non stochastic case the model allows for
(past/known) abrupt breaks.

\subsubsection{First Moments}

We turn our attention to a consideration of the time series properties of
the TV-AR($2$)-AGARCH($1,1$) process. Let the triplet $(\Omega ,\{
\tciFourier _{t},t\in 
\mathbb{Z}
\},P)$\ denote a complete probability space with a filtration, $\{
\tciFourier _{t}\}$. $L_{p}$ stands for the space of $P$-equivalence classes
of finite complex random variables with finite $p$-order. Finally, $%
H=L_{2}(\Omega ,\tciFourier _{t},P)$\ stands for a Hilbert space of random
variables with finite first and second moments. Assuming that the drift and
the two AR time varying coefficients $\phi _{i}(t)$, $i=0,1,2$,\ are non
stochastic and taking the conditional expectation of eq. (\ref{TVAR(2)SOL})
with respect to the $\sigma $ field $\tciFourier _{t-k}$ yields the $k$%
-step-ahead optimal (in $L_{2}$-sense) linear predictor of $y_{t}$\ 
\begin{equation}
\mathbb{E}(y_{t}\left \vert \tciFourier _{t-k}\right. )=\sum_{r=0}^{k-1}\xi
_{t,r}\phi _{0}(t-r)+\xi _{t,k}y_{t-k}+\phi _{2}(t-k+1)\xi _{t,k-1}y_{t-k-1}.
\label{CE-TVAR(2)}
\end{equation}

In addition, the forecast error for the above $k$-step-ahead predictor, $%
\mathbb{FE}(y_{t}\left \vert \tciFourier _{t-k}\right. )=y_{t}-\mathbb{E}%
[y_{t}\left \vert \tciFourier _{t-k}\right. ]$, is given by%
\begin{equation}
\mathbb{FE}(y_{t}\left \vert \tciFourier _{t-k}\right. )=\sum_{r=0}^{k-1}\xi
_{t,r}\varepsilon _{t-r},  \label{FE-TVAR(2)}
\end{equation}%
which is a linear combination of $k$ error terms from time $t-k+1$\ to time $%
t$, where the time varying coefficients, $\xi _{t,r}$, are (for $r\geq 2$)
the determinants of an $r\times r$ tridiagonal matrix ($\Phi _{t,r}$); each
nonzero variable diagonal of this matrix consists of the AR time varying
coefficients $\phi _{i}(\mathfrak{\cdot })$, $i=1,2$\ from time $t-r+i$\ to $%
t$.

The Assumption below provides conditions that are used to obtain the
equivalent of the Wold decomposition for non-stationary time varying
processes with non stochastic coefficients.

Assumption 1. $\sum_{r=0}^{k}\xi _{t,r}\phi _{0}(t-r)$ as $k\rightarrow
\infty $ converges for all $t$, and $\sum_{r=0}^{\infty }\sup_{t}(\xi
_{t,r}^{2}\mathbb{\sigma }_{t-r}^{2})<M<\infty $, $M\in \mathbb{Z}^{+}$.

The challenge we face is that in the time varying models we can not invert
the AR polynomial because of the presence of time dependent coefficients. We
overcome this difficulty and formulate a type of time varying Wold
decomposition theorem (see also Singh and Peiris, 1987; Kowalski and Szynal,
1991).

Under Assumption 1 the model in eq. (\ref{TVAR(2)}) with non stochastic
coefficients admits a second-order MA($\infty $) representation: 
\begin{equation}
y_{t}\overset{L_{2}}{=}\lim_{k\rightarrow \infty }y_{t,k}^{par}\overset{L_{2}%
}{=}\sum_{r=0}^{\infty }\xi _{t,r}[\phi _{0}(t-r)+\varepsilon _{t-r}],
\label{TVAR(2)IMA}
\end{equation}%
which is a unique solution of the TV-AR($2$)-AGARCH($1,1$) model (\ref%
{TVAR(2)}). In other words $y_{t}$ is decomposed into a non random part and
a zero mean random part. In particular, the time dependent first moment:%
%
\begin{equation}
\mathbb{E}(y_{t})=\lim_{k\rightarrow \infty }\mathbb{E}(y_{t}\left\vert
\tciFourier _{t-k}\right. )=\sum_{r=0}^{\infty }\xi _{t,r}\phi _{0}(t-r)
\label{E-TVAR(2)}
\end{equation}%
is the non random part of $y_{t}$ while $\lim_{k\rightarrow \infty }\mathbb{%
FE}(y_{t}\left\vert \tciFourier _{t-k}\right. )=\sum_{r=0}^{\infty }\xi
_{t,r}\varepsilon _{t-r}$ is the zero mean random part.%

The time varying expected value of $y_{t}$ is an infinite sum of the time
varying drifts where the time varying coefficients are expressed as
determinants of continuant matrices (the $\xi ${\footnotesize s}).

\subsubsection{Second Moments}

The current Section and Section \ref{SubsubsecErrorVariance} below discusses
the second-order properties of the TV-AR($2$)-AGARCH($1,1$) model. Next we
state the results for the second moment structure.

The mean square error%
\begin{equation}
\mathbb{V}ar[\mathbb{FE(}y_{t}\left \vert \tciFourier _{t-k}\right.
)]=\sum_{r=0}^{k-1}\xi _{t,r}^{2}\mathbb{\sigma }_{t-r}^{2}
\label{VFE-TVAR(2)}
\end{equation}%
is a linear combination of $k$\ variances from time $t-k+1$\ to
time $t$, with time varying coefficients (the squared $\xi ${\footnotesize s}%
). 

Moreover, under Assumption 1 the second time varying unconditional moment of 
$y_{t}$ exists and it is given by%
\begin{equation}
\mathbb{E}(y_{t}^{2})=[\mathbb{E}(y_{t})]^{2}+\sum_{r=0}^{\infty }\xi
_{t,r}^{2}\sigma _{t-r}^{2},  \label{VAR-TVAR(2)}
\end{equation}%
which is an infinite sum of the time varying unconditional variances of the
errors, $\sigma _{t-r}^{2}$, (see Section \ref{SubsubsecErrorVariance}
below) with time varying `coefficients' or weights (the squared values of
the $\xi ${\footnotesize s}).

In addition, the time varying autocovariance function $\gamma _{t,k}$ is
given by 
\begin{eqnarray}
\gamma _{t,k} &=&\mathbb{C}ov(y_{t},y_{t-k})=\sum_{r=0}^{\infty }\xi
_{t,k+r}\xi _{t-k,r}\sigma _{t-k-r}^{2}  \label{Covar-TVAR(2)} \\
&=&\xi _{t,k}\mathbb{V}ar(y_{t-k})+\phi _{2}(t-k+1)\xi _{t,k-1}\mathbb{C}%
ov(y_{t-k},y_{t-k-1}),  \notag
\end{eqnarray}%
where the second equality follows from the MA($\infty $) representation of $%
y_{t}$ in eq. (\ref{TVAR(2)IMA}), and the third one from the general
solution in eq. (\ref{TVAR(2)SOL}), and%
\begin{equation*}
\mathbb{C}ov(y_{t-k},y_{t-k-1})=\sum_{r=0}^{\infty }\xi _{t-k,r+1}\xi
_{t-k-1,r}\sigma _{t-k-1-r}^{2}.
\end{equation*}

For any fixed $t$, $\lim_{k\rightarrow \infty }\gamma _{t,k}\rightarrow 0$
when $\lim_{k\rightarrow \infty }\xi _{t,k}=0$ $\forall $ $t$. For the
process with $n$ abrupt breaks in eq. (\ref{ASBAR(2)}) $\xi _{t,k}$\ is
given by eq. (\ref{KSIABAR(2)}).\footnote{%
Estimating the time varying parameters of forecasting models is beyond the
scope of this paper (see Elliott and Timmermann, 2008, for an excellent
survey on forecasting methodologies available to the applied economist).}

FIGURE 1 HERE

As an illustrative example Figure 1 shows the autocorrelations (ACR) of an
AR($1$) model with three breaks and homoscedastic/independent innovations.
The left graph in Panel B shows the first order ACR, $\mathbb{C}or(y_{t-i}$, 
$y_{t-i-1})$, for an AR($1$) model with three breaks at times $t-k_{1}(=100)$%
, $t-k_{2}(=120)$ and $t-k_{3}(=140)$, and autoregressive coefficients $\phi
_{1,1}=0.98$, $\phi _{1,2}=0.80$, $\phi _{1,3}=0.70$, and $\phi _{1,4}=0.90$%
. The first part of the graph shows the ACR when $i<k_{1}=100$, that is,
when $y_{t-i}$ is after all three breaks: $t-i>t-k_{1}$ (the construction of
the autocorrelations is based on eq. (\ref{Covar-TVAR(2)})).\footnote{%
The details are available upon request. See also the Additional Appendix,
which is available on the personal webpage of the first author:
http://www.mkaranasos.com/PublicationsB.htm
\par
\bigskip
\par
{}} As $i$ increases, that is, as we are going back in time, the first order
ACR decrease at an increasing rate. The second part of the graph shows the
ACR when $k_{1}\leq i\leq k_{2}-1$, that is, when $y_{t-i}$ is between the
first and the second break. The third part of the graph shows the ACR when $%
k_{2}\leq i\leq k_{3}-1$. The ACR increase since after the third break the
autoregressive coefficient increases from $0.70$ to $0.90$. Finally, for $%
i\geq k_{3}$, the first order ACR are not affected by the three breaks and
therefore are equal to $\phi _{1,4}=0.90$, whereas when $i\rightarrow
-\infty $, the ACR converge to $\phi _{1,1}=0.98$.

Moreover, the right graph in Panel C shows the seventh order ACR ($y_{t-i}$, 
$y_{t-i-7}$) for an AR($1$) model with three breaks at times $t-k_{1}(=100)$%
, $t-k_{2}(=121)$ and $t-k_{3}(=142)$, autoregressive coefficients $\phi
_{1,1}=0.60$, $\phi _{1,2}=1.20$, $\phi _{1,3}=0.80$, and $\phi _{1,4}=0.92$
and homoscedastic/independent innovations. The second part of the graph
shows the ACR when $i\leq k_{1}-1$ and $k_{1}+1\leq i+7\leq k_{2}$. The
fourth part of the graph shows the ACR when $k_{1}$ $\leq i\leq k_{2}-1$ and 
$k_{2}+1\leq i+7\leq k_{3}$. The sixth part of the graph shows the ACR when $%
k_{2}\leq i\leq k_{3}-1$ and $k_{3}+1\leq i+7$. Notice that when $i\leq
k_{1}-1$ or $k_{2}\leq i\leq k_{3}-1$ the seventh order ACR increase with $i$
whereas when $k_{1}$ $\leq i\leq k_{2}-1$ they decrease as $i$ increases.
Finally, for $i\geq k_{3}$, the ACR are equal to $\phi _{1,4}^{7}=0.56$,
whereas when $i\rightarrow -\infty $, the ACR converge to $\phi
_{1,1}^{7}=0.03$.

\subsection{The Conditional Variance\label{SubsecVariance}}

In order to simplify the description of the analysis of this Section we will
introduce the following notation. As before $t$ represents the present time
and $k$ the prediction horizon. We define the bivariate function $\varsigma :%
\mathbb{Z}\times \mathbb{Z}^{+}\longmapsto \mathbb{R}$ by%
\begin{equation}
\varsigma _{t,k}=\dprod \nolimits_{j=0}^{k-1}c(t-j),  \label{zets}
\end{equation}%
coupled with the initial values $\varsigma _{t,0}=1$, and $\varsigma
_{t,-1}=0$ where $c(\mathfrak{\cdot })$ has been defined above (see eq. (\ref%
{TV-ARMA(2,2)})). In other words $\varsigma _{t,1}=c(t)$, and $\varsigma
_{t,k}$ for $k\geq 2$ is a product of $k$ terms which consist of the time
varying coefficients $c(\mathfrak{\cdot )}$ from time $t-k+1$ to time $t$.
For the GARCH process with $m$ abrupt breaks in eq. (\ref{ABR-AGARCH(1,1)})
we have%
\begin{equation}
\varsigma _{t,k}=\dprod \nolimits_{\ell =0}^{m}c_{\ell +1}^{\kappa _{\ell
+1}-\kappa _{\ell }}.  \label{zetsAbrupt}
\end{equation}

Next, we define 
\begin{equation}
g_{t,r+1}=\varsigma _{t,r}\alpha ^{\ast }(t-r),\text{ }r\geq 0,  \label{g}
\end{equation}%
where $\alpha ^{\ast }(t)$ has been defined in eq. (\ref{TVGARCH}). Notice
that when $r=0$, $g_{t,1}=\alpha ^{\ast }(t)$, since $\varsigma _{t,0}=1.$

Since the TV-AGARCH($1,1$) model can be interpreted as a `TV-ARMA($1,1$)'
process, it follows directly from the results in Section \ref%
{SubsecMeanTheorConsid} that the general solution of eq. (\ref{TV-ARMA(2,2)}%
) with free constant (initial condition value) $h_{t-k}$, is given by%
\begin{equation}
h_{t,k}^{gen}=h_{t,k}^{hom}+h_{t,k}^{par}=\varsigma
_{t,k}h_{t-k}+\sum_{r=0}^{k-1}\varsigma _{t,r}\omega
(t-r)+\sum_{r=1}^{k}g_{t,r}v_{t-r},  \label{TVGARCH(2,2)SOl}
\end{equation}%
where $\varsigma _{t,r}$ and $g_{t,r}$ have been defined in eqs. (\ref{zets}%
) and (\ref{g}) respectively. In the above expression $h_{t}^{gen}$\ is
decomposed into two parts: the $h_{t,k}^{hom}$\ part, which is written in
terms of the free constant ($h_{t-k}$); and the $h_{t,k}^{par}$\ part, which
contains the\ time varying drift terms, $\omega (\mathfrak{\cdot })$, and
the uncorrelated terms ($v${\footnotesize s}). Notice that in eq. (\ref%
{TVGARCH(2,2)SOl}) $h_{t,k}^{gen}$ is expressed in terms of diagonal
determinants (the $\varsigma ${\footnotesize s} and therefore the $g$%
{\footnotesize s}).

Next consider the case of a GARCH($1,1$) model with constant coefficients.
Since for this model $\alpha (t)\triangleq a$, and $c(t)\triangleq
c\triangleq \alpha +b$, for all $t$, then $\varsigma _{t,k}$ reduces to $%
\emph{c}^{k}$ and $g_{t,k}$ becomes $\emph{c}^{k-1}a$, for $k\in \mathbb{Z}%
^{+}$ (see, for example, Karanasos, 1999).

\subsubsection{Time Varying Unconditional Variances\label%
{SubsubsecErrorVariance}}

In this Section in order to provide a thorough description of the TV-AGARCH($%
1,1$) process given by eq. (\ref{TVGARCH}) we derive, first its multistep
ahead predictor, the associated forecast error and the mean square error,
and second, the first unconditional moment of this process (the second
unconditional moment and the covariance structure are available upon
request).

The $k$-step-ahead predictor of $h_{t}$, $\mathbb{E}(h_{t}\left \vert 
\mathcal{F}_{t-k-1}\right. )$, is readily seen to be\footnote{%
For the issue of temporal aggregation and a discussion of the wider class of
weak GARCH processes see Bollerslev and Ghysels (1996) and Ghysels and
Osborn (2001, pp. 195-197).}%
\begin{equation}
\mathbb{E}(h_{t}\left \vert \mathcal{F}_{t-k-1}\right. )=\sum_{r=0}^{k-1}%
\overline{\varsigma }_{t,r}\omega (t-r)+\overline{\varsigma }_{t,k}h_{t-k},
\label{OPRTVGARCH(2,2)}
\end{equation}%
where, for $r\geq 1$, $\overline{\varsigma }_{t,r}=\mathbb{E}(\varsigma
_{t,r})$.\footnote{$\mathbb{E}(\varsigma _{t,r})=\mathbb{E[}\dprod
\nolimits_{j=0}^{r-1}c(t-j)]=\dprod \nolimits_{j=0}^{r-1}\overline{c}(t-j)$
with $\overline{c}(t)\triangleq \mathbb{E}[c(t)]=\alpha (t)+\beta (t)+\frac{%
\gamma (t)}{2}$. For the process with $m$ abrupt breaks: $\mathbb{E}%
(\varsigma _{t,r})=\dprod \nolimits_{\ell =0}^{m}\overline{c}_{\ell
+1}^{\kappa _{\ell +1}-\kappa _{\ell }}$.} In addition, the forecast error
for the above $k$-step-ahead predictor, $\mathbb{FE}(h_{t}\left \vert 
\mathcal{F}_{t-k-1}\right. )$, is given by%
\begin{equation}
\mathbb{FE}(h_{t}\left \vert \mathcal{F}_{t-k-1}\right.
)=\sum_{r=1}^{k}g_{t,r}v_{t-r}.  \label{FETVGARCH(2,2)}
\end{equation}

Notice that this predictor is expressed in terms of $k$ uncorrelated terms
(the $v${\footnotesize s}) from time $t-k$ to time $t-1$, where the
`coefficients' have the form of diagonal determinants (the $\varsigma $%
{\footnotesize s}). The mean square error is given by%
\begin{equation}
\mathbb{V}ar(h_{t}\left \vert \mathcal{F}_{t-k-1}\right. )=\mathbb{V}ar[%
\mathbb{FE}(h_{t}\left \vert \mathcal{F}_{t-k-1}\right. )]=\widetilde{%
\varkappa }\sum_{r=1}^{k}\overline{g}_{t,r}^{2}\mathbb{E}(h_{t-r}^{2}),
\label{MSETVGARCH(2,2)}
\end{equation}%
where $\overline{g}_{t,r}^{2}=\mathbb{E}(g_{t,r}^{2})$ for $r\geq 1$.%
\footnote{$\mathbb{E}(g_{t,r+1}^{2})=\mathbb{E}(\varsigma _{t,r}^{2})[\alpha
^{2}(t-r)+\gamma ^{2}(t-r)/2+\alpha (t-r)\gamma (t-r)]$ and, for $r\geq 1$, $%
\mathbb{E}(\varsigma _{t,r}^{2})=\dprod \nolimits_{j=0}^{r-1}\mathbb{E[}%
c^{2}(t-j)]$, with $\mathbb{E[}c^{2}(t)]=[\alpha (t)+\beta (t)]^{2}+\gamma
^{2}(t)/2+[\alpha (t)+\beta (t)]\gamma (t)$.} This is expressed in terms of $%
k$ second moments, $\mathbb{E}(h_{t-r}^{2})$, from time $t-k$ to time $t-1$,
where the coefficients are the expectations of the squared coefficients of
the multistep ahead predictor multiplied by $\widetilde{\varkappa }$.
Moreover, the definition of the uncorrelated term $v_{t}$ implies that $%
\mathbb{E}(\varepsilon _{t}^{2}\left \vert \mathcal{F}_{t-k-1}\right. )=%
\mathbb{E}(h_{t}\left \vert \mathcal{F}_{t-k-1}\right. )$, $\mathbb{FE}%
(\varepsilon _{t}^{2}\left \vert \mathcal{F}_{t-k-1}\right. )=v_{t}+\mathbb{%
FE}(h_{t}\left \vert \mathcal{F}_{t-k-1}\right. )$. The associated mean
squared error is given by $\mathbb{V}ar[\mathbb{FE}(\varepsilon
_{t}^{2}\left \vert \mathcal{F}_{t-k-1}\right. )]=\widetilde{\varkappa }%
\mathbb{E}(h_{t}^{2})+\mathbb{V}ar[\mathbb{FE}(h_{t}\left \vert \mathcal{F}%
_{t-k-1}\right. )]=\widetilde{\varkappa }\sum_{r=0}^{k}\overline{g}_{t,r}^{2}%
\mathbb{E}(h_{t-r}^{2})$.

Next to obtain the first unconditional moment of $h_{t}$, for all $t$, we
impose the conditions that: $\sum_{r=0}^{k}\overline{\varsigma }_{t,r}\omega
(t-r)$ as $k\rightarrow \infty $ is positive and converges, and 
\begin{equation}
\widetilde{\varkappa }\dsum \nolimits_{r=1}^{\infty }\sup \nolimits_{t}[%
\overline{g}_{t,r}^{2}\mathbb{E}(h_{t-r}^{2})]<M<\infty ,\text{ }M\in 
\mathbb{Z}^{+},  \label{CG1}
\end{equation}%
which guarantees that, for all $t$, the model in eq. (\ref{TV-ARMA(2,2)})
admits the second-order MA($\infty $) representation:%
%
\begin{equation}
h_{t,\infty }^{gen}=\lim_{k\rightarrow \infty }h_{t,k}^{par}\overset{L_{2}}{=%
}\sum_{r=0}^{\infty }\overline{\varsigma }_{t,r}\omega
(t-r)+\sum_{r=1}^{\infty }g_{t,r}v_{t-r},  \label{GARCH-MA-INFIN}
\end{equation}%
which is a unique solution of the TV-AGARCH($1,1$) model in eq. (\ref%
{TVGARCH}). The above result states that $\{h_{t,k}^{par}$, $t\in \mathbb{Z}%
^{+}\}$ (defined in eq. (\ref{TVGARCH(2,2)SOl})) $L_{2}$ converges as $%
k\rightarrow \infty $ if and only if $\sum_{r=0}^{k}\overline{\varsigma }%
_{t,r}\omega (t-r)$ as $k\rightarrow \infty $ converges and $%
\sum_{r=1}^{k}g_{t,r}v_{t-r}$ converges a.s., and thus under the
aforementioned conditions $h_{t,\infty }^{gen}\overset{L_{2}}{=}%
\lim_{k\rightarrow \infty }h_{t,k}^{par}$ satisfies eq. (\ref%
{TVGARCH(2,2)SOl}).

Moreover, the first time varying unconditional moment of $h_{t}$, $\mathbb{E}%
(h_{t})=\sigma _{t}^{2}$, is the limit of the ($k+1$)-step-ahead predictor
of $h_{t}$, $\mathbb{E}(h_{t}\left \vert \mathcal{F}_{t-k-1}\right. )$, as $%
k\rightarrow \infty $: 
\begin{equation}
\mathbb{E}(h_{t})=\lim_{k\rightarrow \infty }\mathbb{E}(h_{t}\left \vert 
\mathcal{F}_{t-k-1}\right. )=\sum_{r=0}^{\infty }\overline{\varsigma }%
_{t,r}\omega (t-r).  \label{FirstMomentCondVar}
\end{equation}

Notice that the first moment is time varying. The expected value of the
conditional variance, that is the unconditional variance of the error, is an
infinite sum of the time varying drifts where the coefficients (the $%
\overline{\varsigma }${\footnotesize s}) are expressed as expectations of
diagonal determinants. Finally, for the process with $m$ abrupt breaks in
eq. (\ref{ABR-AGARCH(1,1)}), for $i\leq \kappa _{1}$ we have (if and only if 
$\overline{c}_{m+1}<1$):%
\begin{equation}
\mathbb{E}(h_{t-i})=\frac{1-\overline{c}_{1}^{\kappa _{1}-i}}{1-\overline{c}%
_{1}}\omega _{1}+\dsum\limits_{\ell =2}^{m}\widetilde{c}_{\ell }\frac{1-%
\overline{c}_{_{\ell }}^{\kappa _{_{\ell }}-\kappa _{_{\ell }-1}}}{1-%
\overline{c}_{_{\ell }}}\omega _{_{\ell }}+\widetilde{c}_{m+1}\frac{1}{1-%
\overline{c}_{_{m+1}}}\omega _{_{m+1}},  \label{FirstMomentCondVarAbrupt}
\end{equation}%
with%
\begin{equation*}
\widetilde{c}_{\ell }=\overline{c}_{1}^{\kappa
_{1}-i}\dprod\nolimits_{j=2}^{\ell -1}(\overline{c}_{j}^{\kappa _{j}-\kappa
_{j-1}}),
\end{equation*}%
where we use the convention $\dprod\nolimits_{r=i}^{j}(\cdot )=1$ for $j<i$,
and the $\omega ${\footnotesize s }and the $c${\footnotesize s }are defined
in eqs. (\ref{TVGARCH}) and (\ref{TV-ARMA(2,2)}) respectively. Notice that
if and only if $\overline{c}_{1}<1$ the above expression as $i\rightarrow
-\infty $ becomes: $\mathbb{E}(h_{t-i})=\frac{\omega _{1}}{1-\overline{c}_{1}%
}$ since $\widetilde{c}_{\ell }=\overline{c}_{1}^{\kappa _{1}-i}=0$ for all $%
\ell $. Finally, when $i>\kappa _{m}$, that is when we are before all the
breaks, then if and only if $\overline{c}_{m+1}<1$: $\mathbb{E}(h_{t-i})=%
\frac{\omega _{m+1}}{1-\overline{c}_{m+1}}$.

\section{Methodology and Data}

This Section outlines the methodology we have employed to study the
different properties of the stochastic processes during the various
financial crises and offers an overview of the data employed. First, we
describe the univariate models we have estimated. Then we mention the break
identification method which we have adopted.

\subsection{Univariate Modelling}

Let stock returns be denoted by\textbf{\ }$r_{t}=(\log p_{t}-\log
p_{t-1})\times 100$, where $p_{t}$ is the stock price index, and define its
mean equation as:

\begin{equation}
r_{t}=\mu +\phi _{1}r_{t-1}+\phi _{2}r_{t-2}+\varepsilon _{t}\text{,}
\end{equation}%
where $\varepsilon _{t}\mid \mathcal{F}_{t-1}\sim N(0,$ $h_{t})$, that is
the innovation is conditionally normal with zero mean and variance $h_{t}$.%
\footnote{%
Since mainly structural breaks in the variance are found statistically
significant (see Section $5.1$ below) we do not include any dummies in the
mean. Moreover, low order AR specifications capture the serial correlation
in stock returns.} Next, the dynamic structure of the conditional variance
is specified as an AGARCH($1,1$) process of Glosten et al. (1993) (the
asymmetric power ARCH could also be employed, as in Karanasos and Kim,
2006). In order to examine the impact of the breaks on the persistence of
the conditional variances, the following equation is specified as follows:

\begin{eqnarray}
h_{t} &=&\omega +\sum_{i=1}^{7}\omega _{i}D_{i}+\alpha \varepsilon
_{t-1}^{2}+\sum_{i=1}^{7}\alpha _{i}D_{i}\varepsilon _{t-1}^{2}+\gamma
S_{t-1}^{-}\varepsilon _{t-1}^{2}+\sum_{i=1}^{7}\gamma
_{i}D_{i}S_{t-1}^{-}\varepsilon _{t-1}^{2}  \notag \\
&&+\beta h_{t-1}+\sum_{i=1}^{7}\beta _{i}D_{i}h_{t-1},  \label{GJRB}
\end{eqnarray}%
where $S_{t-1}^{-}=1$ if $e_{t-1}<0,$ and $0$ otherwise. Note that failure
to reject $H_{0}:$ $\gamma =0$ and $\gamma _{i}=0$, $i=1,\ldots ,7$, implies
that the conditional variance follows a symmetric GARCH($1,1$) process.
Furthermore, the second order conditions require that $\overline{c}<1$ and $%
\overline{c}+\dsum \limits_{i=1}^{7}\overline{c}_{i}<1$.\footnote{$\overline{%
c}\triangleq \alpha +\beta +\frac{\gamma }{2}$ and $\overline{c}%
_{i}\triangleq \alpha _{i}+\beta _{i}+\gamma _{i}/2$.} The breakdates $%
i=1,....,7$ are given in Table $1$, and $D_{i}$ are dummy variables defined
as $0$ in the period before each break and one after the break.\footnote{%
The relation between the parameters{\footnotesize \ }in{\footnotesize \ }eq.
(\ref{GJRB}) and the ones{\footnotesize \ }in eq. (\ref{ABR-AGARCH(1,1)}) is
given by, i.e., for the $\omega ${\footnotesize s:} $\omega
+\sum_{_{i}=1}^{m+1-\ell }\omega _{_{i}}=\omega _{\ell }$, $\ell =1,\ldots
,m+1$, where the $\omega ${\footnotesize s} in the right hand side are the
ones in eq. (\ref{ABR-AGARCH(1,1)}).} We also consider a simple GARCH($1,1$)
model which allows the dynamics of the conditional variances to switch
across positive and negative stock returns. This is given by

\begin{equation}
h_{t}=\omega +\omega ^{-}D_{t-1}^{-}+\alpha \varepsilon _{t-1}^{2}+\alpha
^{-}D_{t-1}^{-}\varepsilon _{t-1}^{2}+\beta h_{t-1}+\beta
^{-}D_{t-1}^{-}h_{t-1}.  \label{GARCHPN}
\end{equation}%
where $D_{t-1}^{-}=1$\ if $r_{t-1}<0$, $0$ otherwise\textbf{.}\footnote{%
We estimate another specification with $\alpha ^{+}D_{t-1}^{+},$ $\beta
^{+}D_{t-1}^{+},$ and $\omega ^{+}D_{t-1}^{+}$, instead of $\alpha
^{-}D_{t-1}^{-},$ $\beta ^{-}D_{t-1}^{-},$ and $\omega ^{-}D_{t-1}^{-}$,
where $D_{t-1}^{+}=1$\ if $r_{t-1}>0$, $0$ otherwise. The results (not
reported) are very similar.}\textbf{\ }This is an example of a TV-AGARCH
model with stochastic coefficients.

\subsection{Data and Breaks Overview}

We use daily data that span the period 1-1-1988 30-6-2010 for the stock
market indices, obtained from Thomson DataStream. To account for the
possibility of breaks in the mean and/or volatility dynamics we use a set of
non-parametric data-driven methods to identify the number and timing of the
potential structural breaks. In particular, we adopt the two-stage
Nominating-Awarding procedure of Karoglou (2010) to identify breaks that
might be associated either to structural changes in the mean and/or
volatility dynamics or to latent non-linearities that may manifest
themselves as dramatic changes in the mean and/or volatility dynamics and
might bias our analysis.\footnote{%
The details of the two stages in the Nominating-Awarding procedure and a
summary of the statistical properties of stock market returns are available
upon request.} Alternatively, we could choose the break points by employing
the methodologies in Kim and Kon (1999), Bai and Perron (2003) and Lavielle
and Moulines (2000) (see, for example, Karanasos and Kartsaklas, 2009, and
Campos et al., 2012).

\section{Empirical Analysis}

This Section presents the empirical results we obtain from the different
econometric tools. First, we present the breaks that we have identified and
discuss the possible economic events that may be associated with them. Then
we focus on the stock market returns and condition our analysis based on
these breaks to discuss first the findings from the univariate modelling and
then from the bivariate one (presented in Section 6).

\subsection{Estimated Breaks}

After applying the Nominating-Awarding procedure on stock market returns we
find that the stochastic behaviour of all indices yields about three to
seven breaks during the sample period, roughly one every two to four years
on average. The predominant feature of the underlying segments is that
mainly changes in variance are found statistically significant. Finally,
there are several breakdates that are either identical in all series or very
close to one another, which apparently signify economic events with a global
impact.

It appears that dates for the extraordinary events of the Asian financial
crisis of 1997, the global financial crisis of 2007--08 and the European
sovereign-debt crisis that followed are clearly identified in all stock
return series with very little or no variability (see Table $1$). Other less
spectacular events, such as the Russian financial crisis of 1998, the
Japanese asset price bubble of 1986-1991 or the UK's withdrawal from the
European Exchange Rate Mechanism (ERM), can also be associated with the
breakdates that have been identified in some series.\footnote{%
A detailed account of the possible associations that can be drawn between
each breakdate\textbf{\ }for stock returns and a major economic event that
took place at or around the breakdate period either in the world or in each
respective economy is available upon request, as is a summary of the
descriptive statistics of each segment.}

\subsection{Univariate Results}

The quasi-maximum likelihood estimates of the AGARCH($1,1$) model allowing
the drifts (the $\omega ${\footnotesize s}) as well as the `dynamics of the
conditional variance' (the $\alpha ${\footnotesize s}, $\beta $%
{\footnotesize s} and $\gamma ${\footnotesize s}) to switch across the
considered breaks, as in eq. (\ref{GJRB}), are reported in Table $2$.%
\footnote{%
The quasi-maximum likelihood estimates of the standard AGARCH($1,1$) model
are available upon request. The results of the symmetric GARCH ($1,1$) model
allowing the dynamics of the conditional variance to switch across the
considered breaks are reported in Paraskevopoulos et al. (2013).} The
estimated models are shown to be well-specified: there is no linear or
nonlinear dependence in the residuals in all cases, at the $5\%$ level. Note
that the insignificant parameters are excluded. The impact of the breaks on
the $\omega $ is insignificant in all eight cases. However, there exists a
significant impact of the breaks on the `dynamic structure of the
conditional variance' for all stock returns (irrespective of whether a
symmetric GARCH($1,1$) or an AGARCH ($1,1$) model is considered). \ More
specifically, while the ARCH parameter shows time varying features across a
single break in the cases of S\&P and DAX, for CAC and Hang Seng it is
shifted across two breaks and for STRAITS it is shifted across three breaks
(see the $\alpha _{i}$ coefficients). With regard to the GARCH parameter,
CAC and NIKKEI show time varying parameters for only one break, but S\&P,
TSE, and FTSE across two breaks. Furthermore, the GARCH parameter shows a
time varying pattern across three breaks in the case of DAX and across five
breaks in the\ case of STRAITS.

Interestingly, the asymmetry parameter also displays significant time
variation over the considered breaks. Specifically, the TSE, DAX, and Hang
Seng cases are significantly shifted for one break, whereas S\&P, CAC, and
FTSE show a time varying pattern across three breaks, and STRAITS for two
breaks (see the $\gamma _{i}$ coefficients in Table $2$). Furthermore, the
results are shown to be robust by considering the dynamics of a GARCH($1,1$)
process to switch across positive and negative stock returns (see Table $3$%
). Clearly, the ARCH and GARCH parameters show time dependence across
positive and negative returns in all cases (see the $\alpha ^{-}$, and $%
\beta ^{-}$ coefficients).

Overall, Table $4$ shows that the persistence of the conditional variances
of stock returns varies over the considered breaks in all cases by
considering the AGARCH ($1,1$) models. The persistence is measured by%
{\footnotesize \ }$\overline{c}_{\ell }=\alpha _{\ell }+\beta _{\ell
}+\gamma _{\ell }/2$, $\ell =1,\ldots ,m+1$ (these are the $\overline{c}$%
{\footnotesize s }used in eq. (\ref{FirstMomentCondVarAbrupt}) as well),
and, for example,{\footnotesize \ }$\beta _{\ell }=\underset{\text{Eq. (\ref%
{GJRB})}}{\underbrace{\beta +\dsum\nolimits_{i=1}^{m+1-\ell }\beta _{i}}}$.%
\footnote{%
The plot of the time varying-piecewise persistence of the conditional
variances of stock returns against the persistence generated from the
standard AGARCH($1,1$) models is available upon request.}

The cases which are shown to have been impacted strongly by the breaks are
those of TSE, DAX, Hang Seng, NIKKEI and STRAITS. In particular, the
persistence of the conditional variance of TSE increases from $0.93$\ to $%
0.98$\ after the break in 1996, remains $0.98$ during the recent financial
crisis\ and then increases to near unity after the European sovereign-debt
crisis. With regard to the persistence of the conditional variance of DAX,
it appears to be unaffected by German reunification, its highest value is $%
0.98$ during the Asian financial crisis, its lowest value is $0.94$ after
the break \ associated with the announcement of the \euro $18$bn tax cuts
plan in Germany ($17/06/03$), it increases to $0.97$ on the onset of the
recent financial crisis and remains there during the sovereign-debt crisis.
The results also suggest that the persistence of the conditional variance of
Hang Seng declines from $0.97$ to $0.92$ (its lowest value) after the
savings deposits were removed in July 2001, increases to $0.99$ during the
recent financial crisis in 2007/2008, and finally it declines to $0.94$
after the European sovereign-debt crisis. Furthermore, the corresponding
persistence of STRAITS increases from $0.87$\ to near unity ($0.99$) after
the Asian financial crisis. However, such persistence declines after the
break in June 2000 to $0.91$, remains the same through the unexpected
economic recession in Singapore in 2001 before bounding back to $0.97$\ at
the onset of the global financial crisis, and then exhibits\ a sharp decline
to $0.88$ during the European sovereign-debt crisis. Surprisingly, the
persistence of the conditional variance of NIKKEI increases from $0.90$\ to
approximately $0.98$\ during the asset price bubble in Japan over the period
1986-1991 and remains unaffected afterwards. For example, the impact of the
Asian financial crisis as well as that of the recent financial crisis are
shown to be limited, which may be due to the fact that Japan has been immune
to such crises.\textbf{\ }

The persistence of the conditional variances by allowing the GARCH ($1,1$)
process to switch across positive and negative returns also shows a time
varying pattern (see Table $5$). In particular, it is shown that the
persistence of the conditional variances stemming from positive returns is
lower than those of the negative counterparts. More specifically, positive
returns are shown to lower the persistence of the conditional variances in
most of the cases to around $0.90$ whereas the persistence of the negative
returns is close to unity ($0.99$).

FIGURE 2 HERE

Figure 2 shows the estimated time varying unconditional variances for the
eight stock index returns. For the S\&P the first part of the graph shows
the unconditional variances when $i<k_{1}$, that is, when $h_{t-i}$ is after
all three breaks ($t-k_{3}$(=03/97), $t-k_{2}$(=09/08) and $t-k_{1}$%
(=03/09)) (we construct the time varying unconditional variances using the
formula in eq. (\ref{FirstMomentCondVarAbrupt})). When $i\rightarrow -\infty 
$, the unconditional variances converge to $\omega /(1-\overline{c}%
_{1})=0.001/(1-0.990)=0.100$. As $i$ increases, that is, as we are going
back in time, the unconditional variances increase at an increasing rate.
The second part of the graph shows the unconditional variances when $%
k_{1}\leq i\leq k_{2}-1$, that is, when $h_{t-i}$ is between the first and
the second break. Higher values of $i$ are associated with lower
unconditional variances. When $i=k_{1}$, the unconditional variance is $[(1-%
\overline{c}_{2}^{k_{2}-k_{1}})/(1-\overline{c}_{2})+\overline{c}%
_{2}^{k_{2}-k_{1}}(1-\overline{c}_{3}^{k_{3}-k_{2}})/(1-\overline{c}_{3})+%
\overline{c}_{2}^{k_{2}-k_{1}}\overline{c}_{3}^{k_{3}-k_{2}}/(1-\overline{c}%
_{4})]\omega =0.228$ (see eq. (\ref{FirstMomentCondVarAbrupt}) and the $%
\overline{c}${\footnotesize s }in the first column of Table $4$). The third
part of the graph shows the unconditional variances when $k_{2}\leq i\leq
k_{3}-1$. When $i=k_{2}$, the unconditional variance is $[(1-\overline{c}%
_{3}^{k_{3}-k_{2}})/(1-\overline{c}_{3})+\overline{c}_{3}^{k_{3}-k_{2}}/(1-%
\overline{c}_{4})]\omega =0.105$. Finally, for $i\geq k_{3}$, the
unconditional variances are not affected by the three breaks and therefore
are equal to $\omega /(1-\overline{c}_{4})=0.061$.

Similarly, for the DAX the first part of the graph shows the unconditional
variances when $i<k_{1}$, that is, when $h_{t-i}$ is after all three breaks (%
$t-k_{3}$(=07/97), $t-k_{2}$(=06/03) and $t-k_{3}$(=01/08)). When $%
i\rightarrow -\infty $, the unconditional variances converge to $\omega /(1-%
\overline{c}_{1})=0.011/(1-0.976)=0.458$. As $i$ increases, that is, as we
are going back in time, the unconditional variances decrease at an
increasing rate. The second part of the graph shows the unconditional
variances when $k_{1}\leq i\leq k_{2}-1$ ($\mathbb{E}(h_{t-k_{1}})=0.177$).
Higher values of $i$ are associated with higher unconditional variances. The
third part of the graph shows the unconditional variances when $k_{2}\leq
i\leq k_{3}-1$. They are decreasing with $i$. Finally, for $i\geq k_{3}$,
the unconditional variances are not affected by the three breaks and
therefore are equal to $\omega /(1-\overline{c}_{4})=0.222$.

For the NIKKEI the first part of the graph shows the unconditional variances
when $i<k_{1}$, that is, when $h_{t-i}$ is after the only break ($t-k_{1}$%
(=02/90)). When $i\rightarrow -\infty $, the unconditional variances
converge to $\omega /(1-\overline{c}_{1})=0.326$. As $i$ increases the
unconditional variances decrease at an increasing rate. In addition, for $%
i\geq k_{1}$, the unconditional variances are not affected by the break and
therefore are equal to $\omega /(1-\overline{c}_{2})=0.068$.

Finally, STRAITS exhibits the highest number of breaks, that is six. The
first part of the graph shows the unconditional variances when $i<k_{1}$,
that is, when $h_{t-i}$ is after all six breaks ($t-k_{6}$(=08/91), $t-k_{5}$%
(=08/97), $t-k_{4}$(=06/00), $t-k_{3}$(=07/07), $t-k_{2}$(=05/09), $t-k_{1}$%
(=08/09)). As $i$ increases, that is, as we are going back in time, the
unconditional variances increase at an increasing rate. When $i\rightarrow
-\infty $, the unconditional variances converge to $\omega /(1-\overline{c}%
_{1})=0.157$. The second part of the graph shows the unconditional variances
when $k_{1}\leq i\leq k_{2}-1$. Higher values of $i$ are associated with
higher unconditional variances. The third part of the graph shows the
unconditional variances when $k_{2}\leq i\leq k_{3}-1$. They are decreasing
with $i$. For the fourth and sixth part the unconditional variances increase
with $i$ whereas for the fifth part they decrease with $i$. Finally, for $%
i\geq k_{6}$, the unconditional variances are not affected by the six breaks
and therefore are equal to $\omega /(1-\overline{c}_{7})=0.238$.

\section{Bivariate Models}

In this Section we use a bivariate extension of the univariate formulation
of Section 4.. In particular, we use a bivariate model to simultaneously
estimate the conditional means, variances, and covariances of stock returns.
Let $\mathbf{y}_{t}=(y_{1,t},y_{2,t})^{\prime }$ represent the $2\times 1$
vector with the two returns. $\mathcal{F}_{t-1}=\sigma (\mathbf{y}_{t-1},%
\mathbf{y}_{t-2},\ldots )$ is the filtration generated by the information
available up through time $t-1$. We estimate the following bivariate AR($2$%
)-AGARCH($1,1$) model 
\begin{equation}
\mathbf{y}_{t}=\mathbf{\mu }+\mathbf{\Phi }_{1}\mathbf{y}_{t-1}+\mathbf{\Phi 
}_{2}\mathbf{y}_{t-2}+\mathbf{\varepsilon }_{t},  \label{Mean}
\end{equation}%
where $\mathbf{\mu }=[\mu _{i}]_{i=1,2}$ is a $2\times 1$ vector of drifts
and $\mathbf{\Phi }_{l}=[\phi _{ij}^{(l)}]_{i,j=1,2}$, $l=1,2$, is a $%
2\times 2$ matrix of autoregressive parameters. We assume that the roots of $%
\left\vert \mathbf{I}-\sum_{l=1}^{2}\mathbf{\Phi }_{l}L^{l}\right\vert $
(where $\mathbf{I}$ is the $2\times 2$ identity matrix) lie outside the unit
circle.

Let $\mathbf{h}_{t}=(h_{1,t}$,$h_{2,t})^{\prime }$ denote the $2\times 1$
vector of $\mathcal{F}_{t-1}$ measurable conditional variances. The residual
vector is defined as $\mathbf{\varepsilon }_{t}=(\varepsilon _{1,t}$,$%
\varepsilon _{2,t})^{\prime }=[\mathbf{e}_{t}\odot \mathbf{q}_{t}^{\wedge
-1/2}]\odot \mathbf{h}_{t}^{\wedge 1/2}$, where the symbols $\odot $ and $%
^{\wedge }$ denote the Hadamard product and the elementwise exponentiation
respectively. The stochastic vector $\mathbf{e}_{t}=(e_{1,t}$,$%
e_{2,t})^{\prime }$ is assumed to be independently and identically
distributed ($i.i.d.$) with mean zero, conditional variance vector $\mathbf{q%
}_{t}=(q_{11,t},q_{22,t})^{\prime }$, and $2\times 2$ conditional
correlation matrix\textbf{\ }$\mathbf{R}_{t}=diag\{\mathbf{Q}_{t}\}^{-1/2}%
\mathbf{Q}_{t}diag\{\mathbf{Q}_{t}\}^{-1/2}$\textbf{\ }with diagonal
elements equal to one and off-diagonal elements absolutely less than one%
\textbf{. }A typical element of $\mathbf{R}_{t}$ takes the form $\rho
_{ij,t}=q_{ij,t\text{ }}/\sqrt{q_{ii,t\text{ }}q_{jj,t\text{ }}}$\ for $%
i,j=1,2$.\textbf{\ }The conditional covariance matrix\textbf{\ }$\mathbf{Q}%
_{t}=[q_{ij,t\text{ }}]_{i,j=1,2}$ is specified as in Engle (2002)

\begin{eqnarray}
\mathbf{Q}_{t} &=&(1-\alpha _{D}-\beta _{D}\mathbf{)}\bar{Q}+\alpha _{D}%
\mathbf{e}_{t-1}\mathbf{e}_{t-1}^{\prime }+\beta _{D}\mathbf{Q}_{t-1},
\label{DCC} \\
&&  \notag
\end{eqnarray}%
where\textbf{\ }$\bar{Q}$\textbf{\ }is the unconditional covariance matrix of%
\textbf{\ }$\mathbf{e}_{t}$\textbf{, }and $\alpha _{D}$\ and $\beta _{D}$\
are non-negative scalars fulfilling $\alpha _{D}$\ $+$\ $\beta _{D}<1.$

Following Conrad and Karanasos (2010) and Rittler (2012), we impose the
UEDCC-AGARCH($1,1$) structure on the conditional variances (multivariate
fractionally integrated APARCH models could also be used, as in Conrad et
al., 2011 or \ Karanasos et al., 2014), and we also amend it by allowing the
shock and volatility spillovers parameters to be time varying:%
\begin{eqnarray}
\mathbf{h}_{t} &=&\mathbf{\omega }+\mathbf{A}^{\ast }\mathbf{\varepsilon }%
_{t-1}^{\wedge 2}+\sum_{l=1}^{n}\mathbf{A}_{l}D_{l}\mathbf{\varepsilon }%
_{t-1}^{\wedge 2}+\mathbf{B}\mathbf{h}_{t-1}+\sum_{l=1}^{n}\mathbf{B}%
_{l}D_{l}\mathbf{h}_{t-1},  \label{variance} \\
&&  \notag
\end{eqnarray}%
where $\mathbf{\omega }=[\omega _{i}]_{i=1,2}$, $\mathbf{A}=[\alpha
_{ij}]_{i,j=1,2}$, $\mathbf{B}=[\beta _{ij}]_{i,j=1,2}$; $\mathbf{A}_{l}$, $%
l=1,\ldots ,n$ (and $n=0,1,\ldots ,7$) is a cross diagonal matrix with
nonzero elements $\alpha _{ij}^{l}$, $i,j=1,2$, $i\neq j$, and $\mathbf{B}%
_{l}$, is a cross diagonal matrix with nonzero elements $\beta _{ij}^{l}$, $%
i,j=1,2$, $i\neq j$; $\mathbf{A}^{\ast }=\mathbf{A+\Gamma S}_{t-1}$, $%
\mathbf{\Gamma }$ is a diagonal matrix with elements $\gamma _{ii}$, $i=1,2$%
, and $\mathbf{S}_{t-1}$ is a diagonal matrix with elements $S_{i,t-1}^{-}=1$
if $e_{i,t-1}<0$, $0$ otherwise. The model without the breaks for the shock
and volatility spillovers, that is $\mathbf{h}_{t}=\mathbf{\omega }+\mathbf{A%
}^{\ast }\mathbf{\varepsilon }_{t-1}^{\wedge 2}+\mathbf{B}\mathbf{h}_{t-1}$,
is minimal in the sense of Jeantheau (1998, Definition 3.3) and invertible
(see Assumption~2 in Conrad and Karanasos, 2010). The invertibility
condition implies that the inverse roots of $\left\vert \mathbf{I}-\mathbf{B}%
L\right\vert $, denoted by $\varphi _{1}$ and $\varphi _{2}$, lie inside the
unit circle. 
Following Conrad and Karanasos (2010) we also impose the four conditions
which are necessary and sufficient for $\mathbf{h}_{t}\succ 0$ for all $t$:
(i) $(1-b_{22})\omega _{1}+b_{12}\omega _{2}>0$ and $(1-b_{11})\omega
_{2}+b_{21}\omega _{1}>0$, (ii) $\varphi _{1}$ is real and $\varphi
_{1}>|\varphi _{2}|$, (iii) $\mathbf{A}^{\ast }\succeq 0$ and (iv) $[\mathbf{%
B}-$max$(\varphi _{2},0)\mathbf{I}]\mathbf{A}^{\ast }\succ 0$, where the
symbol $\succ $ denotes the elementwise inequality operator. Note that these
constraints do not place any \emph{a priori} restrictions on the signs of
the coefficients in the $\mathbf{B}$ matrix. In particular, these
constraints imply that negative volatility spillovers are possible. When the
conditional correlations are constant, the model reduces to the UECCC-GARCH($%
1,1$) specification of Conrad and Karanasos (2010).

Finally, we also amend the UEDCC-AGARCH($1,1$) model by allowing shocks and
volatility spillovers to vary across positive and negative returns: 
\begin{equation*}
\mathbf{h}_{t}=\mathbf{\omega }+\mathbf{A}^{\ast }\mathbf{\varepsilon }%
_{t-1}^{\wedge 2}+\mathbf{B}^{\ast }\mathbf{h}_{t-1},
\end{equation*}%
where $\mathbf{A}^{\ast }=\mathbf{A+\Gamma S}_{t-1}+\mathbf{A}^{-}\mathbf{D}%
_{t-1}^{-}$ and $\mathbf{B}^{\ast }=\mathbf{B}+\mathbf{B}^{+}\mathbf{D}%
_{t-1}^{+}$; $\mathbf{A}^{-}$($\mathbf{B}^{+}$) is a cross diagonal matrix
with nonzero elements $\alpha _{ij}^{-}$($\beta _{ij}^{+}$), $i,j=1,2$, $%
i\neq j$; $\mathbf{D}_{t}^{-}$($\mathbf{D}_{t}^{+}$) are $2\times 1$ vectors
with elements $d_{it}^{-}$($d_{it}^{+}$), $i=1,2$, where $d_{it}^{-}$($%
d_{it}^{+}$) is one if $r_{jt}<0$ ($r_{jt}>0$) and zero otherwise, $j=1,2$, $%
j\neq i$.

\subsection{Bivariate Results}

\textit{Example 1: FTSE-DAX}

\bigskip

Table $6$\ reports the results of the UEDCC-AGARCH($1,1$) model between the
returns on FTSE and DAX allowing shock and volatility spillover parameters
to shift across the breaks in order to analyze the time varying volatility
transmission structure between the two variables\textbf{.}\footnote{%
For an application on the returns of commodity metal futures see Karanasos
et al. (2013).} As is evident from Table $6$, the results suggest the
existence of strong conditional heteroscedasticity in the two variables. The
ARCH as well as the asymmetry parameters of the two variables are positive
and significant, indicating the existence of asymmetric responses in the two
variables. In addition, rejection of the model with constant conditional
correlation, using Tse's (2000) test, indicates the time varying conditional
correlation between the two financial markets.\textbf{\ }Figure $3$ displays
the evolution of the time varying conditional correlation between the two
variables over the sample period.

Furthermore, the results suggest that there is evidence of shock spillovers
as well as negative volatility spillovers from DAX to FTSE (the $\alpha
_{12} $ and $\beta _{12}$ coefficients are significant at the $1\%$ and $%
10\% $ levels, respectively).\footnote{%
The results for the conventional UEDCC-GARCH($1,1$) process are available
upon request. For this model the stationarity condition of Engle (2002) is
satisfied over time.} With regard to the impact of the breaks on the
volatility transmission structure, it is shown that both shock and
volatility spillovers between the two variables change over time. The most
significant changes include the impact of the fourth break in DAX
(15/01/2008), which corresponds to the global financial crisis, in which it
shifts the shock spillovers parameter from DAX to FTSE (the $\alpha
_{12}^{4} $ coefficient is significant at the $1\%$ level). Also, volatility
spillovers from DAX to FTSE are shown to be shifted after the second
(21/07/1997) and the third break (17/06/2003), corresponding to the Asian
financial crisis and the announcement of the \euro $18$bn German tax cuts
plan, respectively (see the $\beta _{12}^{2}$ and $\beta _{12}^{3}$
coefficients in Table $6$). 

These results are consistent with the time varying conditional correlations.
The average time varying conditional correlation for the period before the
break 15/01/2008 is $0.58$ compared to the period after the break of $0.89$.
This also applies for the break 21/07/1997 (17/06/2003) with an average time
varying correlation of $0.43$ ($0.52$) for the period before the break and $%
0.75$ ($0.82$) for the period after the break. Overall these findings are
indicative of the existence of contagion between DAX\ and FTSE during the
turbulent periods of the two financial crises.

FIGURE 3 HERE

Another way to look at the structure of the volatility spillovers between
DAX and FTSE is to allow volatility (and shock) spillover parameters to
shift across two regimes of stock returns: positive (increases in the stock
market) and negative (declines in the stock market) returns. The results,
displayed in Table $7$, suggest that declines in each market generate shock
spillovers to the other (the coefficients $\alpha _{12}^{-}$ and $\alpha
_{21}^{-}$ are positive and significant), whilst increases in each market
generate negative volatility spillovers to the other (the coefficients $%
\beta _{12}^{+}$ and $\beta _{21}^{+}$ are negative and significant). 

\bigskip

\textit{Example 2: NIKKEI-Hang Seng}

\bigskip \ 

Next, we consider the structure of the volatility spillovers between the
returns on NIKKEI and Hang Seng\textbf{\ }to provide an example about the
dynamic linkages between the Asian financial markets. The estimated
bivariate model, reported in Table $8$, suggests the existence of strong
conditional heteroscedasticity. There is evidence of asymmetric effects of
the two variables as the ARCH and asymmetry parameters (the $\alpha $ and
the $\gamma $ coefficients) are positive and significant. Furthermore, the
model with constant conditional correlation is rejected according to Tse's
(2000) test, hence the correlation between the two variables is time
varying. This is also confirmed by Figure $4$, which shows the evolution of
the time varying correlation between the two variables.

With regard to the linkages between the two variables, the results show the
existence of shock spillovers from Hang Seng\ to NIKKEI after the third
(05/05/2009) and the fourth break (01/12/2009), which correspond to the
different phases of the\textbf{\ }European sovereign-debt crisis. Also,
while Hang Seng generates negative volatility spillovers to NIKKEI\ after
the third break in the former (05/05/2009), there are positive volatility
spillovers from NIKKEI to Hang Seng after the second break (04/01/2008) in
the former, which corresponds to the global financial crisis. These findings
indicate the superiority of the time varying spillover model over the
conventional one. In contrast to the conventional model, allowing for breaks
shows that the two financial markets have been integrated during the global
financial crisis.\footnote{%
The results from the conventional bivariate UEDCC-AGARCH($1,1$) process
indicate that there is no evidence of volatility spillovers between the two
financial markets (they are available upon request). For this model the
stationarity condition of Engle (2002) is fulfilled.}

With regard to the time varying conditional correlations, the average time
varying conditional correlation for the period before the breaks 04/01/2008,
05/05/2009, and 01/12/2009 are respectively $0.40,0.41$, and $0.415$
compared to the period after the breaks of $0.60,0.58,$ and $0.585$,
respectively. These results are consistent with those of volatility
spillovers in which these two types of markets have become more dependent
during the recent financial crisis\textbf{.}

FIGURE 4 HERE

Finally, allowing the volatility spillover structure to shift across two
different regimes, that is, positive and negative returns, also shows the
existence of time varying volatility spillovers between the two variables.
Specifically, the results, displayed in Table $9$, suggest that declines in
NIKKEI generate shock spillovers to Hang Seng (the estimated $\alpha
_{21}^{-}$ coefficient is positive and significant), whilst increases in
NIKKEI generate negative volatility spillovers to Hang Seng (the estimated $%
\beta _{21}^{+}$ coefficient is negative and significant).\textbf{\ }

\section{Summary and Conclusions}

In this paper, we have introduced a platform to examine empirically the link
between financial crises and the principal time series properties of the
underlying series. We have also adopted several models, both univariate and
bivariate, to examine how the mean and volatility dynamics, including the
volatility persistence and volatility spillovers structure of stock market
returns have changed due to the recent financial crises and conditioned our
analysis on non-parametrically identified breaks. Overall, our findings are
consistent with the intuitively familiar albeit empirically hard-to-prove
time varying nature of asset market linkages induced by economic events and
suggest the existence of limited diversification opportunities for
investors, especially during turbulent periods.

In particular, with respect to the mean and volatility dynamics our findings
suggest that in general the financial crises clearly affect more the
(un)conditional variances. Also, the results of the volatility persistence
are clear-cut and suggest that they exhibit substantial time variation. This
time variation applies to all stock market returns irrespective of whether
we allow for structural changes or positive and negative changes in the
underlying market. As far as the direction of this time variation during
financial crises is concerned the jury is still out, but there is little
doubt that the financial crises are the primary driving force behind the
profound changes in the unconditional variances.

Finally, with respect to the existence of dynamic correlations as well as
time varying shock and volatility spillovers our findings are also
conclusive. Specifically, they suggest that in the cases we examine there is
an increase in conditional correlations, occurring at different phases of
the various financial crises, hence providing evidence as to the existence
of contagion during these periods. Such a finding is comparable to those of
other studies using only conditional correlation analysis to examine the
existence of contagion during the various financial crises. The results also
suggest the existence of regime dependent volatility spillovers in all cases
we examine by using two regimes of returns, positive and negative. Given
that this is to our knowledge the first attempt to take into account the
joint effect of dynamic correlations, volatility spillovers and structural
breaks in the mean and/or volatility dynamics, these findings are of
particular interest to those seeking refuge from financial crises.

\bigskip

\textbf{REFERENCES}

\bigskip

Asgharian, H., Nossman, M., 2011. Risk contagion among international stock
markets. Journal of International Money and Finance 30, 22-38.

Baele, L., 2005. Volatility spillover effects in European equity markets.
Journal of Financial and Quantitative Analysis 40, 373--401.

Bai, J., Perron, P., 2003. Computation and analysis of multiple structural
change models. Journal of Applied Econometrics 18, 1--22.

Baillie, R.T., Morana, C., 2009. \ Modelling long memory and structural
breaks in conditional variances: An adaptive FIGARCH approach. Journal of
Economic Dynamics and Control 33, 1577-1592.

Bauwens, L., Dufays, A., Rombouts, J. V. K., 2014. Marginal likelihood for
Markov-switching and change-point GARCH models. Journal of Econometrics 178,
508-522.

Bollerslev, T., Ghysels, E., 1996. Periodic autoregressive conditional
heteroscedasticity. Journal of Business \& Economic Statistics 14, 139-151.

Campos, N.F., Karanasos, M.G., Tan, B., 2012. Two to tangle: Financial
development, political instability and economic growth in Argentina. Journal
of Banking and Finance 36, 290-304.

Caporale, G. M., Cipollini, A., Spagnolo, N., 2005. Testing for contagion: A
conditional correlation analysis. Journal of Empirical Finance 12, 476--489.

Caporale, G. M., Hunter, J., Menla Ali, F., 2014. On the linkages between
stock prices and exchange rates:\ Evidence from the banking crisis of
2007-2010. International Review of Financial Analysis, forthcoming.

Conrad, C., Karanasos, M., 2010. Negative volatility spillovers in the
unrestricted ECCC-GARCH model. Econometric Theory, 26, 838-862.

Conrad, C., Karanasos, M., 2013. Modeling the link between US inflation and
output: the importance of the uncertainty channel. University of Heidelberg,
Department of Economics, Discussion Paper No. 507. (update of Discussion
Paper No.475).

Conrad, C., Karanasos, M., Zeng, N., 2010. The link between macroeconomic
performance and variability in the UK. Economics Letters 106, 154-157.

Conrad, C., Karanasos, M., Zeng, N., 2011. Multivariate fractionally
integrated APARCH modeling of stock market volatility: A multi-country
study. Journal of Empirical Finance 18, 147-159.

Elliott, G., Timmermann, A., 2008. Economic forecasting. Journal of Economic
Literature 46, 3-56.

Engle, R. F., 2002. Dynamic conditional correlation: A simple class of
multivariate generalized autoregressive conditional heteroskedasticity
models. Journal of Business \& Economic Statistics 20, 339-350.

Engle, R .F., Rangel, J.G., 2008. The spline-GARCH model for low-frequency
volatility and its global macroeconomic causes. The Review of Financial
Studies 21, 1187-1222.

Frijns, B. Lehnert, T., Zwinkels, R.C.J., 2011. Modeling structural changes
in the volatility process. Journal of Empirical Finance 18, 522-532.

Francq, C., Zako\"{\i}an, J-M., 2010. GARCH Models. Structure, Statistical
Inference and Financial Applications. Chichester, West Sussex, UK. Wiley.

Glosten, L. R., Jagannathan, R., Runkle, D. E., 1993. On the relation
between the expected value and the volatility of nominal excess return on
stocks. The Journal of Finance 46, 1779-1801.

Ghysels, E., Osborn, D. R., 2001. The Econometric Analysis of Seasonal Time
Series. Cambridge University Press.

Granger, C.W.J., 2007. Forecasting - looking back and forward: paper to
celebrate the 50th anniversary of the Econometrics Institute at the Erasmus
University, Rotterdam. Journal of Econometrics 138, 3-13.

Granger, C.W.J., 2008. Non-linear models: where do we go next - time varying
parameter models?, Studies in Nonlinear Dynamics \& Econometric\textit{s}
12, 1-9.

Hosking, J. R. M., 1981. Equivalent Forms of the Multivariate Portmanteau
Statistic. Journal of Royal Statistical Society-Series B 43, 261-262.

Jeantheau, T., 1998. Strong consistency of estimators for multivariate ARCH
models. Econometric Theory 14, 70--86.

Karanasos, M., 1998. A new method for obtaining the autocovariance of an
ARMA model: An exact form solution. Econometric Theory 14, 622-640.

Karanasos, M., 1999. The second moment and the autocovariance function of
the squared errors of the GARCH model. Journal of Econometrics 90, 63-76.

Karanasos, M., 2001. Prediction in ARMA models with GARCH in mean effects.
Journal of Time Series Analysis 22, 555-576.

Karanasos, M., Kartsaklas, A. 2009. Dual long-memory, structural breaks and
the link between turnover and the range-based volatility. Journal of
Empirical Finance 16, 838-851.

Karanasos, M., Kim., J., 2006. A re-examination of the asymmetric power ARCH
model. Journal of Empirical Finance 13, 113-128.

Karanasos, M., Menla Ali, F., Margaronis, Z. P., and Nath, R.B., 2013.
Modelling time varying volatility spillovers and conditional correlations
across commodity metal futures. Unpublished paper.

Karanasos, M., Zeng, N., 2013. Conditional heteroskedasticity in
macroeconomics data: UK inflation, output growth and their uncertainties, in
Handbook of Research Methods and Applications in Empirical Macroeconomics,
edited by Nigar Hashimzade and Michael A. Thornton. Cheltenham, UK, Edward
Elgar Publishing.

Karanasos, M., Paraskevopoulos, A., and Dafnos, S., 2014a. The fundamental
properties of time varying AR models with non stochastic coefficients.
arXiv:1403.3359.

Karanasos, M., Paraskevopoulos, A., and Dafnos, S., 2014b. A univariate time
varying analysis of periodic ARMA processes. arXiv:1403.4803.

Karanasos, M., Yfanti, S., and Karoglou, M., 2014. Multivariate FIAPARCH
modelling of financial markets with dynamic correlations in times of crisis.
Unpublished Paper.

Karoglou, M., 2010. Breaking down the non-normality of stock returns. The
European Journal of Finance 16, 79--95.

Kim, D., Kon, S. J., 1999. Structural change and time dependence in models
of stock returns. Journal of Empirical Finance 6, 283-308.

Kowalski, A., Szynal, D., 1991. On a characterization of optimal predictors
for nonstationary ARMA processes. Stochastic Processes and their
Applications 37, 71-80.

Lavielle, M., Moulines, E., 2000. Least-squares estimation of an unknown
number of shifts in a time series. Journal of Time Series Analysis 21,
33--59.

Margaronis, Z. P., Nath, R. B., Karanasos, M., and Menla Ali., F., 2013. The
significance of rollover in commodity returns using PARCH models.
Unpublished paper.

Morana, C., Beltratti, A., 2004. Structural change and long-range dependence
in volatility of exchange rates: either, neither or both? Journal of
Empirical Finance 11, 629-658.

Paraskevopoulos, A.G., 2012. The infinite Gauss-Jordan elimination on
row-finite $\omega \times \omega $\TEXTsymbol{\backslash} matrices. arXiv:
1201.2950.

Paraskevopoulos, A.G., 2014. The solution of row-finite linear systems with
the infinite Gauss-Jordan elimination. arXiv: 1403.2624.

Paraskevopoulos, A., Karanasos, M., 2013. Closed form solutions for linear
difference equations with time dependent coefficients. Unpublished paper.

Paraskevopoulos, A., Karanasos, M., Dafnos, S., 2013.\textbf{\ }A unified
theory for time varying models: foundations with applications in the
presence of breaks and heteroscedasticity (and some results on companion and
Hessenberg matrices). Unpublished Paper.

Pesaran, M, H., Pettenuzzo, D., Timmermann, A., 2006. Forecasting time
series subject to multiple structural breaks. The Review of Economic Studies
73, 1057-1084.

Pesaran, M. H., Timmermann, A., 2005. Small sample properties of forecasts
from autoregressive models under structural breaks. Journal of Econometrics
129, 183-217.

Rittler, D., 2012. Price discovery and volatility spillovers in the European
Union emissions trading scheme: A high-frequency analysis. Journal of
Banking \& Finance 36, 774-785.

Rodriguez, J. C., 2007. Measuring financial contagion: A copula approach.
Journal of Empirical Finance 14, 401--423.

Singh, N., Peiris, M.S., 1987. A note on the properties of some
nonstationary ARMA processes. Stochastic Processes and their Applications,
24, 151-155.

Tse, Y. K., 2000. A test for constant correlations in a multivariate GARCH
model. Journal of Econometrics, 98, 107-127.

\newpage

\singlespacing

\begin{tabular}{ccccccccc}
\multicolumn{9}{c}{\textrm{Table 1}} \\ \hline\hline
\multicolumn{9}{c}{\textrm{The break points (Stock Returns)}} \\ 
\textrm{Break} & $\mathrm{S\&P}$ & $\mathrm{TSE}$ & $\mathrm{CAC}$ & $%
\mathrm{DAX}$ & $\mathrm{FTSE}$ & $\mathrm{Hang\ Seng}$ & $\mathrm{NIKKEI}$
& $\mathrm{STRAITS}$ \\ \cline{2-9}
$1$ & $\mathbf{27/03/97}$ & $\mathbf{05/11/96}$ & $\mathbf{17/03/97}$ & $%
27/08/91$ & $\mathbf{22/10/92}$ & $\mathbf{24/10/01}$ & $\mathbf{21/02/90}$
& $\mathbf{26/08/91}$ \\ 
$2$ & $\mathbf{04/09/08}$ & $15/01/08$ & $\mathbf{31/07/98}$ & \underline{$%
\mathbf{21/07/97}$} & $\mathbf{13/07/98}$ & $\mathbf{27/07/07}$ & \underline{%
$04/01/08$} & $\mathbf{28/08/97}$ \\ 
$3$ & $\mathbf{31/03/09}$ & $\mathbf{02/04/09}$ & $\mathbf{15/01/08}$ & 
\underline{$\mathbf{17/06/03}$} & $\mathbf{24/07/07}$ & \underline{$05/05/09$%
} & $03/04/09$ & $\mathbf{06/06/00}$ \\ 
$4$ & $16/07/09$ & $\mathbf{19/08/09}$ & $03/04/09$ & \underline{$\mathbf{%
15/01/08}$} & $06/04/09$ & \underline{$\mathbf{01/12/09}$} &  & $\mathbf{%
26/07/07}$ \\ 
$5$ & $27/04/10$ &  & $27/04/10$ & $03/04/09$ & $27/04/10$ &  &  & $\mathbf{%
28/05/09}$ \\ 
$6$ &  &  &  &  &  &  &  & $\mathbf{25/08/09}$ \\ 
$7$ &  &  &  &  &  &  &  & $28/04/10$ \\ \hline\hline
\multicolumn{9}{l}{{\footnotesize Notes: The dates in bold indicate
breakdates for which, in the univariate estimation (see Table }$%
{\footnotesize 2}${\footnotesize ), at least one}} \\ 
\multicolumn{9}{l}{{\footnotesize dummy variable is significant, i.e, for
the S\&P index for the }${\footnotesize 04/09/08}$ {\footnotesize breakdate} 
$\beta _{2}$ {\footnotesize and }$\gamma _{2}$ {\footnotesize are
significant. The}} \\ 
\multicolumn{9}{l}{{\footnotesize underlined dates indicate breakdates for
which, in the bivariate estimation (see Tables }$6$ {\footnotesize and }$8$%
{\footnotesize ), at least one}} \\ 
\multicolumn{9}{l}{{\footnotesize dummy variable is significant, i.e., for
the NIKKEI-Hang Seng bivariate model, for the }$01/12/09$ {\footnotesize %
breakdate }$\alpha _{12}^{4}$ {\footnotesize is significant.}}%
\end{tabular}

\begin{tabular}{lcccccccc}
\multicolumn{9}{c}{\textrm{Table 2 }} \\ \hline\hline
\multicolumn{9}{c}{\textrm{The estimated univariate AGARCH (1,1) allowing
for breaks in the variance}} \\ 
& \textrm{S\&P} & \textrm{TSE} & \textrm{CAC} & \textrm{DAX} & \textrm{FTSE}
& \textrm{Hang Seng} & \textrm{NIKKEI} & \textrm{STRAITS} \\ \cline{2-9}
$\mu $ & $\underset{(0.004)}{0.012}^{a}$ & $\underset{(0.003)}{0.011}^{a}$ & 
$\underset{(0.006)}{0.010}^{c}$ & $\underset{(0.005)}{0.019}^{a}$ & $%
\underset{(0.004)}{0.009}^{b}$ & $\underset{(0.005)}{0.019}^{a}$ & $\underset%
{(0.005)}{0.006}$ & $\underset{(0.005)}{0.010}^{b}$ \\ 
$\phi _{1}$ &  & $\underset{(0.013)}{0.129}^{a}$ &  &  &  & $\underset{%
(0.014)}{0.079}^{a}$ &  & $\underset{(0.016)}{0.124}^{a}$ \\ 
$\omega $ & $\underset{(0.0002)}{0.001}^{c}$ & $\underset{(0.0007)}{0.003}%
^{a}$ & $\underset{(0.0004)}{0.005}^{a}$ & $\underset{(0.0006)}{0.011}^{a}$
& $\underset{(0.0003)}{0.002}^{a}$ & $\underset{(0.003)}{0.015}^{a}$ & $%
\underset{(0.001)}{0.007}^{a}$ & $\underset{(0.004)}{0.018}^{a}$ \\ 
$\alpha $ & $\underset{(0.006)}{0.018}^{a}$ & $\underset{(0.007)}{0.012}^{c}$
& $\underset{(0.003)}{0.006}^{b}$ & $\underset{(0.006)}{0.031}^{a}$ & $%
\underset{(0.004)}{0.013}^{a}$ & $\underset{(0.007)}{0.039}^{a}$ & $\underset%
{(0.005)}{0.019}^{a}$ & $\underset{(0.010)}{0.018}^{c}$ \\ 
$\alpha _{1}$ & $\underset{(0.008)}{-0.039}^{a}$ &  &  &  &  & $\underset{%
(0.011)}{-0.050}^{a}$ &  & $\underset{(0.013)}{0.059}^{a}$ \\ 
$\alpha _{2}$ &  &  & $\underset{(0.006)}{0.011}^{c}$ &  &  & $\underset{%
(0.014)}{0.068}^{a}$ &  &  \\ 
$\alpha _{3}$ &  &  & $\underset{(0.016)}{-0.044}^{a}$ & $\underset{(0.011)}{%
-0.050}^{a}$ &  &  &  &  \\ 
$\beta $ & $\underset{(0.002)}{0.954}^{a}$ & $\underset{(0.016)}{0.906}^{a}$
& $\underset{(0.003)}{0.936}^{a}$ & $\underset{(0.002)}{0.861}^{a}$ & $%
\underset{(0.001)}{0.952}^{a}$ & $\underset{(0.013)}{0.866}^{a}$ & $\underset%
{(0.026)}{0.820}^{a}$ & $\underset{(0.011)}{0.854}^{a}$ \\ 
$\beta _{1}$ &  &  &  &  & $\underset{(0.002)}{-0.019}^{a}$ &  & $\underset{%
(0.021)}{0.081}^{a}$ & $\underset{(0.029)}{-0.112}^{a}$ \\ 
$\beta _{2}$ & $\underset{(0.009)}{-0.048}^{a}$ &  & $\underset{(0.003)}{%
-0.031}^{a}$ & $\underset{(0.007)}{0.029}^{a}$ & $\underset{(0.006)}{-0.019}%
^{a}$ &  &  & $\underset{(0.029)}{0.115}^{a}$ \\ 
$\beta _{3}$ & $\underset{(0.015)}{0.039}^{a}$ & $\underset{(0.009)}{0.017}%
^{c}$ &  & $\underset{(0.012)}{-0.029}^{b}$ &  &  &  & $\underset{(0.018)}{%
-0.076}^{a}$ \\ 
$\beta _{4}$ &  & $\underset{(0.013)}{-0.025}^{c}$ &  & $\underset{(0.006)}{%
0.038}^{a}$ &  &  &  & $\underset{(0.029)}{0.137}^{a}$ \\ 
$\gamma $ & $\underset{(0.012)}{0.023}^{c}$ & $\underset{(0.009)}{0.028}^{a}$
& $\underset{(0.004)}{0.056}^{a}$ & $\underset{(0.023)}{0.117}^{a}$ & $%
\underset{(0.006)}{0.029}^{a}$ & $\underset{(0.021)}{0.130}^{a}$ & $\underset%
{(0.013)}{0.117}^{a}$ & $\underset{(0.017)}{0.105}^{a}$ \\ 
$\gamma _{1}$ & $\underset{(0.014)}{0.092}^{a}$ & $\underset{(0.023)}{0.097}%
^{a}$ & $\underset{(0.007)}{0.035}^{a}$ &  & $\underset{(0.005)}{0.028}^{a}$
&  &  &  \\ 
$\gamma _{2}$ & $\underset{(0.027)}{0.113}^{a}$ &  & $\underset{(0.009)}{%
0.019}^{b}$ &  & $\underset{(0.016)}{0.055}^{a}$ &  &  &  \\ 
$\gamma _{3}$ & $\underset{(0.029)}{-0.094}^{a}$ &  & $\underset{(0.038)}{%
0.117}^{a}$ & $\underset{(0.043)}{0.075}^{c}$ & $\underset{(0.012)}{0.026}%
^{b}$ &  &  &  \\ 
$LogL$ & $-2921.3$ & $-1837.5$ & $-4374.3$ & $-4469.8$ & $-2904.1$ & $%
-5231.4 $ & $-4764.1$ & $-3957.7$ \\ 
$LB(5)$ & $\underset{[0.138]}{8.343}$ & $\underset{[0.128]}{2.316}$ & $%
\underset{[0.054]}{10.870}$ & $\underset{[0.395]}{5.170}$ & $\underset{%
[0.082]}{9.745}$ & $\underset{[0.231]}{2.928}$ & $\underset{[0.768]}{2.555}$
& $\underset{[0.069]}{3.303}$ \\ 
$LB^{2}(5)$ & $\underset{[0.856]}{1.947}$ & $\underset{[0.979]}{0.759}$ & $%
\underset{[0.556]}{3.953}$ & $\underset{[0.354]}{5.524}$ & $\underset{[0.522]%
}{4.192}$ & $\underset{[0.534]}{4.105}$ & $\underset{[0.109]}{8.992}$ & $%
\underset{[0.897]}{1.635}$ \\ \hline\hline
\multicolumn{9}{l}{{\footnotesize Notes: Robust-standard errors are used in
parentheses. }$LB(5)${\small \ and }$LB^{2}(5)${\footnotesize \ are
Ljung-Box tests for serial correlations}} \\ 
\multicolumn{9}{l}{{\footnotesize of five lags on the standardized and
squared standardized residuals, respectively (}$p$-{\footnotesize values
reported in brackets).}} \\ 
\multicolumn{9}{l}{{\footnotesize Insignificant parameters are excluded. }$%
^{a}${\footnotesize ,} $^{b},$ {\footnotesize and }$^{c}${\footnotesize \
indicate significance at the }${\footnotesize 1}${\footnotesize \%, }$%
{\footnotesize 5}${\footnotesize \%, and }${\footnotesize 10}${\footnotesize %
\% levels, respectively. For }} \\ 
\multicolumn{9}{l}{{\footnotesize the Hang Seng index }$\phi _{3}$ 
{\footnotesize and} $\gamma _{4}${\footnotesize \ are significant, and for
the STRAITS index }$\alpha _{4},\alpha _{6},\beta _{6},\gamma _{5},$ 
{\footnotesize and} $\gamma _{6}$ {\footnotesize are also significant.}}%
\end{tabular}

\begin{tabular}{lllllllll}
\multicolumn{9}{c}{\textrm{Table 3}} \\ \hline\hline
\multicolumn{9}{c}{\textrm{The estimated univariate GARCH (1, 1) models
allowing for different persistence across }} \\ 
\multicolumn{9}{c}{\textrm{positive and negative returns: }$h_{t}=\omega
+\omega ^{-}D_{t-1}^{-}+\alpha \varepsilon _{t-1}^{2}+\alpha
^{-}D_{t-1}^{-}\varepsilon _{t-1}^{2}+\beta h_{t-1}+\beta
^{-}D_{t-1}^{-}h_{t-1}$} \\ 
& \textrm{S\&P} & \textrm{TSE} & \textrm{CAC} & \textrm{DAX} & \textrm{FTSE}
& \textrm{Hang Seng} & \textrm{NIKKEI} & \textrm{STRAITS} \\ \cline{2-9}
$\mu $ & $\underset{(0.005)}{0.036}^{a}$ & $\underset{(0.004)}{0.023}^{a}$ & 
$\underset{(0.007)}{0.044}^{a}$ & $\underset{(0.008)}{0.054}^{a}$ & $%
\underset{(0.004)}{0.032}^{a}$ & $\underset{(0.007)}{0.051}^{a}$ & $\underset%
{(0.007)}{0.034}^{a}$ & $\underset{(0.004)}{0.027}^{a}$ \\ 
$\phi _{1}$ &  & $\underset{(0.012)}{0.114}^{a}$ &  &  &  & $\underset{%
(0.013)}{0.069}^{a}$ &  & $\underset{(0.011)}{0.112}^{a}$ \\ 
$\omega $ & $\underset{(0.0008)}{0.002}^{a}$ & $\underset{(0.0006)}{0.002}%
^{a}$ & $\underset{(0.001)}{0.007}^{a}$ & $\underset{(0.002)}{0.008}^{a}$ & $%
\underset{(0.0005)}{0.002}^{a}$ & $\underset{(0.002)}{0.009}^{a}$ & $%
\underset{(0.0008)}{0.004}^{a}$ & $\underset{(0.002)}{0.006}^{a}$ \\ 
$\alpha $ & $\underset{(0.005)}{0.054}^{a}$ & $\underset{(0.012)}{0.062}^{a}$
& $\underset{(0.008)}{0.070}^{a}$ & $\underset{(0.018)}{0.091}^{a}$ & $%
\underset{(0.006)}{0.066}^{a}$ & $\underset{(0.011)}{0.088}^{a}$ & $\underset%
{(0.008)}{0.065}^{a}$ & $\underset{(0.015)}{0.051}^{a}$ \\ 
$\alpha ^{-}$ &  & $\underset{(0.017)}{0.033^{c}}$ &  &  &  & $\underset{%
(0.020)}{0.033^{c}}$ & $\underset{(0.015)}{0.025^{c}}$ & $\underset{(0.021)}{%
0.104}^{a}$ \\ 
$\beta $ & $\underset{(0.023)}{0.837}^{a}$ & $\underset{(0.027)}{0.861}^{a}$
& $\underset{(0.023)}{0.822}^{a}$ & $\underset{(0.039)}{0.779}^{a}$ & $%
\underset{(0.014)}{0.832}^{a}$ & $\underset{(0.025)}{0.815}^{a}$ & $\underset%
{(0.016)}{0.842}^{a}$ & $\underset{(0.023)}{0.883}^{a}$ \\ 
$\beta ^{-}$ & $\underset{(0.034)}{0.208}^{a}$ & $\underset{(0.024)}{%
0.106^{a}}$ & $\underset{(0.029)}{0.181^{a}}$ & $\underset{(0.043)}{0.233}%
^{a}$ & $\underset{(0.023)}{0.187}^{a}$ & $\underset{(0.037)}{0.141^{a}}$ & $%
\underset{(0.027)}{0.157}^{a}$ &  \\ \hline
$LogL$ & $-2941.2$ & $-1865.7$ & $-4388.4$ & $-4478.8$ & $-2903.4$ & $%
-5260.7 $ & $-4799.1$ & $-4048.6$ \\ 
$LB(5)$ & $\underset{[0.089]}{9.526}$ & $\underset{[0.195]}{1.674}$ & $%
\underset{[0.071]}{3.256}$ & $\underset{[0.484]}{4.464}$ & $\underset{[0.154]%
}{8.031}$ & $\underset{[0.104]}{4.521}$ & $\underset{[0.823]}{2.180}$ & $%
\underset{[0.056]}{3.650}$ \\ 
$LB^{2}(5)$ & $\underset{[0.791]}{2.398}$ & $\underset{[0.989]}{0.573}$ & $%
\underset{[0.515]}{4.237}$ & $\underset{[0.375]}{5.340}$ & $\underset{[0.365]%
}{5.428}$ & $\underset{[0.416]}{4.998}$ & $\underset{[0.134]}{8.430}$ & $%
\underset{[0.793]}{2.385}$ \\ \hline\hline
\multicolumn{9}{l}{{\footnotesize Notes: See notes of Table 2. The }$\phi
_{3}${\footnotesize \ coefficient was significant for the CAC and Hang Seng
indices.}}%
\end{tabular}

\bigskip

\begin{tabular}{ccccccccc}
\multicolumn{9}{c}{\textrm{Table }$\mathrm{4}$} \\ \hline\hline
\multicolumn{9}{c}{\textrm{The persistence of the AGARCH (1,1) models}} \\ 
\hline
&  &  &  &  &  &  &  &  \\ 
\multicolumn{9}{c}{\textrm{The persistence of the standard AGARCH (1,1)
models}} \\ 
& $\mathrm{S\&P}$ & $\mathrm{TSE}$ & $\mathrm{CAC}$ & $\mathrm{DAX}$ & $%
\mathrm{FTSE}$ & $\mathrm{Hang\ Seng}$ & $\mathrm{NIKKEI}$ & $\mathrm{STRAITS%
}$ \\ \cline{2-9}
& $0.986$ & $0.986$ & $0.978$ & $0.979$ & $0.985$ & $0.976$ & $0.990$ & $%
0.990$ \\ 
&  &  &  &  &  &  &  &  \\ 
\multicolumn{9}{c}{\textrm{The persistence of the AGARCH (1,1) allowing for
breaks in the variance}} \\ 
\textrm{Break} & $\mathrm{S\&P}$ & $\mathrm{TSE}$ & $\mathrm{CAC}$ & $%
\mathrm{DAX}$ & $\mathrm{FTSE}$ & $\mathrm{Hang\ Seng}$ & $\mathrm{NIKKEI}$
& $\mathrm{STRAITS}$ \\ \cline{2-9}
$0$ & $(\overline{c}_{4}=)0.983$ & $0.932$ & $0.970$ & $(\overline{c}%
_{4}=)0.950$ & $0.979$ & $0.970$ & $0.897$ & $0.924$ \\ 
$1$ & $(\overline{c}_{3}=)0.990$ & $0.980$ & $0.987$ &  & $0.974$ & $0.920$
& $0.978$ & $0.871$ \\ 
$2$ & $(\overline{c}_{2}=)0.998$ &  & $0.976$ & $(\overline{c}_{3}=)0.979$ & 
$0.982$ & $0.988$ &  & $0.986$ \\ 
$3$ & $(\overline{c}_{1}=)0.990$ & $0.997$ & $0.990$ & $(\overline{c}%
_{2}=)0.937$ & $0.995$ &  &  & $0.910$ \\ 
$4$ &  & $0.972$ &  & $(\overline{c}_{1}=)0.976$ &  & $0.945$ &  & $0.974$
\\ 
$5$ &  &  &  &  &  &  &  & $0.948$ \\ 
$6$ &  &  &  &  &  &  &  & $0.884$ \\ \hline\hline
\multicolumn{9}{l}{{\footnotesize Notes: Break }${\footnotesize 0}$%
{\footnotesize \ covers the period preceding all breaks, while break }$%
{\footnotesize 1}${\footnotesize \ covers the period between break}} \\ 
\multicolumn{9}{l}{${\footnotesize 1}${\footnotesize \ and }${\footnotesize 2%
}$,{\footnotesize \ and break }${\footnotesize 2}${\footnotesize \ covers
the period between break }${\footnotesize 2}${\footnotesize \ and }$%
{\footnotesize 3}${\footnotesize , and so on (see Table }${\footnotesize 1}$%
{\footnotesize \ for the dates of the}} \\ 
\multicolumn{9}{l}{\footnotesize breaks). When the value of the persistence
is left blank for a break, it indicates that such persistence} \\ 
\multicolumn{9}{l}{\footnotesize has not changed during the period covered
by such a break. The persistence is measured by} \\ 
\multicolumn{9}{l}{{\footnotesize \ }$\overline{c}_{\ell }=\alpha _{\ell
}+\beta _{\ell }+\gamma _{\ell }/2$, $\ell =1,\ldots ,m+1$, {\footnotesize %
and, for example, }$\beta _{\ell }=\underset{\text{Eq. (\ref{GJRB})}}{%
\underbrace{\beta +\dsum\nolimits_{i=1}^{m+1-\ell }\beta _{i}}}$. 
{\footnotesize That is }$\overline{c}_{m+1}$} \\ 
\multicolumn{9}{l}{{\footnotesize is the persistence before all breaks, and }%
$\overline{c}_{1}${\footnotesize \ is the persistence after all the breaks.}}%
\end{tabular}

\bigskip

\begin{tabular}{ccccccccc}
\multicolumn{9}{c}{\textrm{Table 5}} \\ \hline\hline
\multicolumn{9}{c}{\textrm{The persistence of the GARCH (1,1) allowing for
different persistence }} \\ 
\multicolumn{9}{c}{\textrm{\ across positive and negative returns}} \\ 
\textrm{Break} & $\mathrm{S\&P}$ & $\mathrm{TSE}$ & $\mathrm{CAC}$ & $%
\mathrm{DAX}$ & $\mathrm{FTSE}$ & $\mathrm{Hang\ Seng}$ & $\mathrm{NIKKEI}$
& $\mathrm{STRAITS}$ \\ \cline{2-9}
$r$ & $0.986$ & $0.986$ & $0.978$ & $0.979$ & $0.985$ & $0.976$ & $0.990$ & $%
0.990$ \\ 
$r^{+}$ & $0.891$ & $0.923$ & $0.892$ & $0.870$ & $0.898$ & $0.903$ & $0.907$
& $0.934$ \\ 
$r^{-}$ & $0.995$ & $0.992$ & $0.982$ & $0.986$ & $0.991$ & $0.990$ & $0.998$
& $0.986$ \\ \hline\hline
\multicolumn{9}{l}{{\footnotesize Notes: }$r$ {\footnotesize denotes} 
{\footnotesize the persistence generated from returns, that is from the
standard AGARCH}} \\ 
\multicolumn{9}{l}{{\footnotesize model whilst }$r^{+}(r^{-})${\normalsize \ 
}{\footnotesize corresponds to the persistence generated from positive
(negative) returns.}}%
\end{tabular}

\bigskip

\begin{tabular}{llllllllllll}
\multicolumn{12}{c}{\textrm{Table 6}} \\ \hline\hline
\multicolumn{12}{c}{\textrm{Coefficient Estimates of Bivariate UEDCC-AGARCH
Models Allowing }} \\ 
\multicolumn{12}{c}{\textrm{\ for Shifts in Volatility Spillovers between
FTSE and DAX}} \\ \hline
\multicolumn{12}{c}{\textrm{Conditional Variance Equation }} \\ 
$\omega _{1}$ &  &  & $\underset{(0.0006)}{0.003}^{a}$ &  &  & $\gamma _{11}$
& $\underset{(0.016)}{0.078}^{a}$ &  &  & $\beta _{12}^{3}$ & $\underset{%
(0.002)}{-0.007}^{a}$ \\ 
$\omega _{2}$ &  &  & $\underset{(0.001)}{0.004}^{a}$ &  &  & $\gamma _{22}$
& $\underset{(0.022)}{0.082}^{a}$ &  &  & $\alpha _{D}$ & $\underset{(0.010)}%
{0.044}^{a}$ \\ 
$\alpha _{11}$ &  &  & $\underset{(0.007)}{0.016}^{b}$ &  &  & $\alpha _{12}$
& $\underset{(0.003)}{0.010}^{a}$ &  &  & $\beta _{D}$ & $\underset{(0.011)}{%
0.952}^{a}$ \\ 
$\alpha _{22}$ &  &  & $\underset{(0.009)}{0.033}^{a}$ &  &  & $\alpha
_{12}^{4}$ & $\underset{(0.004)}{0.011}^{a}$ &  &  &  &  \\ 
$\beta _{11}$ &  &  & $\underset{(0.014)}{0.921}^{a}$ &  &  & $\beta _{12}$
& $\underset{(0.003)}{-0.007}^{c}$ &  &  &  &  \\ 
$\beta _{22}$ &  &  & $\underset{(0.015)}{0.912}^{a}$ &  &  & $\beta
_{12}^{2}$ & $\underset{(0.001)}{0.003}^{a}$ &  &  &  &  \\ 
$LogL$ &  &  & $-5427.03$ &  &  &  &  &  &  &  &  \\ 
$Q(5)$ &  &  & $\underset{[0.110]}{27.970}$ &  &  & $Q^{2}(5)$ & $\underset{%
[0.977]}{9.427}$ &  &  &  &  \\ \hline\hline
\multicolumn{12}{l}{{\footnotesize Notes: Robust-standard errors are used in
parentheses, }${\footnotesize 1}${\footnotesize = FTSE, }${\footnotesize 2}$%
{\footnotesize =DAX. }$Q(5)${\small \ and }$Q^{2}(5)$} \\ 
\multicolumn{12}{l}{\footnotesize are the multivariate Hosking (1981) tests
for serial correlation of five lags on the standardized} \\ 
\multicolumn{12}{l}{{\footnotesize and squared standardized residuals,
respectively (}$p$-{\footnotesize values are reported in brackets).}} \\ 
\multicolumn{12}{l}{$\alpha _{12}(\beta _{12})$ {\footnotesize indicates
shock (volatility) spillovers from DAX to FTSE, while }$\alpha
_{12}^{l}(\beta _{12}^{l})$} \\ 
\multicolumn{12}{l}{{\footnotesize indicates the shift in shock\
(volatility) spillovers for the break }${\footnotesize l}${\footnotesize \
(see Table }${\footnotesize 1}${\footnotesize ) from DAX to FTSE.}} \\ 
\multicolumn{12}{l}{{\footnotesize Insignificant parameters are excluded.}$%
^{a}${\footnotesize \ , }$^{b}${\footnotesize \ and }$^{c}${\footnotesize \
indicate significance at the }${\footnotesize 1\%}${\footnotesize , }$%
{\footnotesize 5\%}${\footnotesize , and}} \\ 
\multicolumn{12}{l}{${\footnotesize 10\%}${\footnotesize \ levels,
respectively. Tse's (2000) test for constant conditional correlation: }$%
20.41 $.}%
\end{tabular}

\begin{tabular}{llllllllllll}
\multicolumn{12}{c}{\textrm{Table 7}} \\ \hline\hline
\multicolumn{12}{c}{\textrm{Coefficient Estimates of Bivariate UEDCC-AGARCH
Models Allowing for Different}} \\ 
\multicolumn{12}{c}{\textrm{Spillovers Across Positive and Negative Returns
(FTSE-DAX) }} \\ \hline
\multicolumn{12}{c}{\textrm{Conditional Variance Equation}} \\ 
$\omega _{1}$ &  &  & $\underset{(0.0005)}{0.002}^{a}$ &  &  & $\gamma _{11}$
& $\underset{(0.012)}{0.058}^{a}$ &  &  & $\alpha _{D}$ & $\underset{(0.010)}%
{0.043}^{a}$ \\ 
$\omega _{2}$ &  &  & $\underset{(0.001)}{0.004}^{a}$ &  &  & $\gamma _{22}$
& $\underset{(0.016)}{0.060}^{a}$ &  &  & $\beta _{D}$ & $\underset{(0.011)}{%
0.954}^{a}$ \\ 
$\alpha _{11}$ &  &  & $\underset{(0.008)}{0.030}^{a}$ &  &  & $\alpha
_{12}^{-}$ & $\underset{(0.005)}{0.019}^{a}$ &  &  &  &  \\ 
$\alpha _{22}$ &  &  & $\underset{(0.008)}{0.027}^{a}$ &  &  & $\beta
_{12}^{+}$ & $\underset{(0.004)}{-0.014}^{a}$ &  &  &  &  \\ 
$\beta _{11}$ &  &  & $\underset{(0.012)}{0.926}^{a}$ &  &  & $\alpha
_{21}^{-}$ & $\underset{(0.015)}{0.042}^{a}$ &  &  &  &  \\ 
$\beta _{22}$ &  &  & $\underset{(0.012)}{0.928}^{a}$ &  &  & $\beta
_{21}^{+}$ & $\underset{(0.016)}{-0.036}^{a}$ &  &  &  &  \\ 
$LogL$ &  &  & $-5430.26$ &  &  &  &  &  &  &  &  \\ 
$Q(5)$ &  &  & $\underset{[0.136]}{26.965}$ &  &  & $Q^{2}(5)$ & $\underset{%
[0.975]}{9.533}$ &  &  &  &  \\ \hline\hline
\multicolumn{12}{l}{{\footnotesize Notes: Robust-standard errors are used in
parentheses, }${\footnotesize 1}${\footnotesize = FTSE, }${\footnotesize 2}$%
{\footnotesize =DAX. }$Q(5)${\small \ and }$Q^{2}(5)$} \\ 
\multicolumn{12}{l}{\footnotesize are the multivariate Hosking (1981) tests
for serial correlation of five lags on the standardized} \\ 
\multicolumn{12}{l}{{\footnotesize and} {\footnotesize squared standardized
residuals, respectively (}$p${\footnotesize -values reported in brackets). }$%
\alpha _{12}^{-}(\beta _{12}^{+})$} \\ 
\multicolumn{12}{l}{\footnotesize indicates the shock (volatility)
spillovers from DAX to FTSE generated by negative(positive)} \\ 
\multicolumn{12}{l}{{\footnotesize returns in DAX.. }$\alpha _{21}^{-}(\beta
_{21}^{+})$\textrm{\ }{\footnotesize reports shock (volatility) spillovers
form FTSE to DAX generated by}} \\ 
\multicolumn{12}{l}{\footnotesize negative(positive) returns in FTSE.
Insignificant parameters are excluded.} \\ 
\multicolumn{12}{l}{$^{a}${\footnotesize \ indicates significance at the 1\%
level.}}%
\end{tabular}

\begin{tabular}{llllllllllll}
\multicolumn{12}{c}{\textrm{Table 8}} \\ \hline\hline
\multicolumn{12}{c}{\textrm{Coefficient Estimates of Bivariate UEDCC-AGARCH
Models Allowing for Shifts}} \\ 
\multicolumn{12}{c}{\textrm{in} \textrm{Volatility} \textrm{Spillovers
between NIKKEI and Hang Seng}} \\ \hline
\multicolumn{12}{c}{\textrm{Conditional Variance Equation}} \\ 
$\omega _{1}$ &  &  & $\underset{(0.0008)}{0.003}^{a}$ &  &  & $\gamma _{11}$
& $\underset{(0.012)}{0.094}^{a}$ &  &  & $\alpha _{D}$ & $\underset{(0.005)}%
{0.015}^{a}$ \\ 
$\omega _{2}$ &  &  & $\underset{(0.002)}{0.009}^{a}$ &  &  & $\gamma _{22}$
& $\underset{(0.021)}{0.081}^{a}$ &  &  & $\beta _{D}$ & $\underset{(0.006)}{%
0.982}^{a}$ \\ 
$\alpha _{11}$ &  &  & $\underset{(0.004)}{0.024}^{a}$ &  &  & $\alpha
_{12}^{3}$ & $\underset{(0.017)}{0.050}^{a}$ &  &  &  &  \\ 
$\alpha _{22}$ &  &  & $\underset{(0.007)}{0.050}^{a}$ &  &  & $\alpha
_{12}^{4}$ & $\underset{(0.011)}{0.025}^{b}$ &  &  &  &  \\ 
$\beta _{11}$ &  &  & $\underset{(0.007)}{0.920}^{a}$ &  &  & $\beta
_{12}^{3}$ & $\underset{(0.015)}{-0.046}^{a}$ &  &  &  &  \\ 
$\beta _{22}$ &  &  & $\underset{(0.015)}{0.885}^{a}$ &  &  & $\beta
_{21}^{2}$ & $\underset{(0.009)}{0.016}^{c}$ &  &  &  &  \\ 
$LogL$ &  &  & $-9413.42$ &  &  & {\small Tse's test}$:$ & $10.10$ &  &  & 
&  \\ 
$Q(5)$ &  &  & $\underset{[0.333]}{22.122}$ &  &  & $Q^{2}(5)$ & $\underset{%
[0.850]}{13.594}$ &  &  &  &  \\ \hline\hline
\multicolumn{12}{l}{{\footnotesize Notes: Robust-standard errors are used in
the parentheses, }${\footnotesize 1}${\footnotesize = NIKKEI, }$%
{\footnotesize 2}${\footnotesize =Hang Seng. }$Q(5)${\small \ and }$Q^{2}(5)$%
} \\ 
\multicolumn{12}{l}{{\footnotesize are the} {\footnotesize multivariate
Hosking (1981) tests for serial correlation of five lags on the standardized}
} \\ 
\multicolumn{12}{l}{{\footnotesize and squared standardized residuals,
respectively (}$p$-{\footnotesize values are reported in brackets).}} \\ 
\multicolumn{12}{l}{{\footnotesize direction. }$\alpha _{12}^{l}(\beta
_{12}^{l})${\footnotesize \ indicates shift in shock (volatility) spillovers
for the break }${\footnotesize l}${\footnotesize \ (see Table 1) from}} \\ 
\multicolumn{12}{l}{{\footnotesize Hang Seng to NIKKEI, whilst }$\beta
_{21}^{l}$ {\footnotesize reports} {\footnotesize the shift in volatility
spillovers for the break }${\footnotesize l}$ {\footnotesize in the}} \\ 
\multicolumn{12}{l}{{\footnotesize reverse direction. Insignificant
parameters are excluded. }$^{a}${\footnotesize \ , }$^{b}${\footnotesize \
and }$^{c}${\footnotesize \ indicate significance at the}} \\ 
\multicolumn{12}{l}{${\footnotesize 1\%}${\footnotesize , }${\footnotesize %
5\%}${\footnotesize , and }${\footnotesize 10\%}${\footnotesize \ levels,
respectively.}}%
\end{tabular}

\bigskip

\begin{tabular}{llllllllllll}
\multicolumn{12}{c}{\textrm{Table 9}} \\ \hline\hline
\multicolumn{12}{c}{\textrm{Coefficient Estimates of Bivariate UEDCC-AGARCH
Models Allowing for Different}} \\ 
\multicolumn{12}{c}{\textrm{Spillovers Across Positive and Negative Returns
(NIKKEI-Hang Seng )}} \\ 
\multicolumn{12}{c}{\textrm{Conditional Variance Equation}} \\ 
$\omega _{1}$ &  &  & $\underset{(0.0009)}{0.003}^{a}$ &  &  & $\beta _{11}$
& $\underset{(0.007)}{0.917}^{a}$ &  &  & $\alpha _{21}^{-}$ & $\underset{%
(0.009)}{0.017}^{a}$ \\ 
$\omega _{2}$ &  &  & $\underset{(0.002)}{0.008}^{a}$ &  &  & $\beta _{22}$
& $\underset{(0.013)}{0.897}^{a}$ &  &  & $\beta _{21}^{+}$ & $\underset{%
(0.008)}{-0.018}^{a}$ \\ 
$\alpha _{11}$ &  &  & $\underset{(0.005)}{0.027}^{a}$ &  &  & $\gamma _{11}$
& $\underset{(0.015)}{0.099}^{a}$ &  &  & $\alpha _{D}$ & $\underset{(0.007)}%
{0.016}^{a}$ \\ 
$\alpha _{22}$ &  &  & $\underset{(0.007)}{0.052}^{a}$ &  &  & $\gamma _{22}$
& $\underset{(0.019)}{0.065}^{a}$ &  &  & $\beta _{D}$ & $\underset{(0.010)}{%
0.980}^{a}$ \\ 
$LogL$ &  &  & $-9414.61$ &  &  &  &  &  &  &  &  \\ 
$Q(5)$ &  &  & $\underset{[0.292]}{22.918}$ &  &  & $Q^{2}(5)$ & $\underset{%
[0.975]}{9.534}$ &  &  &  &  \\ \hline\hline
\multicolumn{12}{l}{{\footnotesize Notes: Robust-standard errors are used in
parentheses, }${\footnotesize 1}${\footnotesize = NIKKEI, }${\footnotesize 2}
${\footnotesize =Hang Seng. }$Q(5)${\small \ and }$Q^{2}(5)$} \\ 
\multicolumn{12}{l}{{\footnotesize are the} {\footnotesize multivariate
Hosking (1981) tests for\ serial correlation of five lags on the standardized%
}} \\ 
\multicolumn{12}{l}{{\footnotesize and squared standardized residuals,
respectively (}$p$-{\footnotesize values are reported in brackets). }$\alpha
_{21}^{-}(\beta _{21}^{+})$} \\ 
\multicolumn{12}{l}{\footnotesize reports shock (volatility) spillovers from
NIKKEI to Hang Seng generated by negative(positive) returns} \\ 
\multicolumn{12}{l}{{\footnotesize in NIKKEI. Insignificant parameters are
excluded. }$^{a}${\footnotesize \ indicates significance at the }$%
{\footnotesize 1\%}$ {\footnotesize level.}}%
\end{tabular}

\newpage

\end{document}